\documentclass[twocolumn,twocolappendix]{aastex701}
\usepackage{amsmath,amstext}
\usepackage{bm}
\usepackage{mathtools}
\usepackage{color}

\graphicspath{{./}{figures/}}
\begin{document}

\title{A Magnetized Black Hole Envelope Model for Little Red Dots}

\correspondingauthor{Shinsuke Takasao}
\author[orcid=0000-0003-3882-3945,sname='Takasao']{Shinsuke Takasao}
\altaffiliation{}
\affiliation{Humanities and Sciences/Museum Careers, Musashino Art University, Tokyo 187-8505, Japan}
\email[show]{stakasao@musabi.ac.jp}  

\author[orcid=0000-0001-9840-4959,sname='Inayoshi']{Kohei Inayoshi}
\altaffiliation{}
\affiliation{Kavli Institute for Astronomy and Astrophysics, Peking University, Beijing 100871, China}
\email[show]{inayoshi@pku.edu.cn}

\begin{abstract}
Recent observations have revealed a unique class of active galactic nuclei (AGNs), termed little red dots (LRDs). These objects are hypothesized to be powered by massive black holes rapidly accreting in dense gaseous environments. Theoretical studies suggest that the circum-nuclear gas can form an optically thick black hole envelope (BHE), whose structure resembles the atmospheres of convective stars near the Hayashi limit. Given that such cool stars typically generate magnetic fields, we propose a dynamical and spectral model for an LRD enshrouded by a magnetized BHE. Assuming spherical free-fall accretion onto a rotating, magnetized BHE, our model accounts for key observational properties of LRDs.
We propose that the Doppler component of broad emission lines originates from plasma clumps co-rotating within the BHE magnetosphere. Including additional broadening due to electron scattering allows the resulting line profile to be fitted by a combination of a Gaussian core and an exponential tail. This model can reproduce Doppler components up to a few thousand ${\rm km~s^{-1}}$. We suggest that conventional black hole mass estimation methods based on the virial relation may yield erroneous results.
Furthermore, our model is consistent with X-ray non-detections in LRDs. We evaluate the X-ray luminosities of two potential sources: the post-shock region of accretion shocks and a magnetically heated corona. We find that these X-ray luminosities are constrained to $\lesssim 10^{41}~{\rm erg~s^{-1}}$ across a wide range of black hole masses ($10^5 M_\odot \lesssim  M_{\rm BH}\lesssim 10^7M_\odot$) and accretion rates, consistent with current upper limits on X-ray emission.
\end{abstract}



\section{Introduction}

The James Webb Space Telescope (JWST) has revolutionized our understanding of the early universe at cosmic dawn through its unprecedented sensitivity to infrared light. The discovery of little red dots (hereafter, LRDs) has provided crucial insights into the emergence of massive black holes and their origins \citep{Matthee_2024,Greene_2024,Kocevski_2025,Labbe_2025,Hviding_2025}. LRDs exhibit unique spectral properties, including V-shaped spectral energy distributions (SEDs) composed of blue ultraviolet and red optical continuum emission, along with prominent Balmer breaks and absorption features \citep{Maiolino_2024_JADES,Lin_2024,Inayoshi_Maiolino_2025,Ji_2025}, and weak hot dust and X-ray emission \citep{Maiolino_2025,Setton_2025b,Perez-Gonzalez_2024,Comastri_2025,Delvecchio_2025}. These characteristics distinguish LRDs from typical active galactic nuclei, suggesting that they represent an early stage of massive black hole formation and growth \citep{Inayoshi_2025a}.

Several theoretical scenarios have been proposed since the discovery of LRDs to explain their nature \citep[][references therein]{Inayoshi_Ho_2025}. Among these, the gas-enshrouded active galactic nuclei (AGN) model successfully accounts for multiple key aspects of their enigmatic properties. In this scenario, the surrounding gas forms an optically thick envelope that reprocesses accretion power into photospheric emission with an effective surface temperature of $T_{\rm eff}\simeq 5000~{\rm K}$ \citep[e.g.,][]{Inayoshi_2025b,Kido2025MNRAS,Naidu_2025,deGraaff_2025b,deGraaff_2025c,Taylor_2025b,Kokorev_2025,Lin_2026,Begelman2026ApJ,Sneppen2026arXiv260118864S}. This framework includes various structural possibilities, such as black hole envelopes (BHEs), quasi-stars, and black hole stars. While these models differ in detail, they share a common feature: an optically thick envelope that behaves like a stellar atmosphere. Consequently, these models self-consistently reproduce the red optical emission and relatively weak infrared output observed in LRDs.

Despite its success in reproducing several observed properties, this configuration leaves the origin of the UV continuum and the broad emission lines unclear. This is primarily because radiation produced by the accreting black hole can hardly escape from the deep interior surrounded by a dense gaseous envelope, suggesting that it is instead reprocessed into optical emission via thermalization. However, if the envelope is clumpy and not perfectly optically thick \citep{Naidu_2025,Sneppen2026arXiv260118864S,Tang2026arXiv260403563T,Ji2026arXiv260403370J}, a fraction of ionizing radiation may escape along low-density sightlines before being thermalized. In this context, a recent study of spectroscopically confirmed LRDs showed that the UV continuum and broad H$\alpha$ luminosities are tightly correlated, following a ratio consistent with that observed in star-forming galaxies \citep{Asada_2026}. Furthermore, the absence or weakness of emission lines from highly ionized species (e.g., \ion{He}{2}, \ion{C}{4}, and [\ion{Ne}{3}]) suggests that the radiation source in LRDs is unlikely to be a typical AGN \citep{Wang_2025b,Tang2025ApJ}; however, \citet{Tang2025ApJ} reported detecting \ion{N}{5} emission, which is difficult to produce via stellar activity.

Hydrogen Balmer lines might alternatively be powered by a population of young stars clustered around the BHE \citep[e.g.,][]{Inayoshi_2025d, Asada_2026} or by ionizing radiation leaking from the envelope interior \citep[e.g.,][]{Lin_2026,Sneppen2026arXiv260118864S}. In either case, if the line-emitting clouds just outside the envelope radiate efficiently, the expected full width at half maximum (FWHM) is $\sim 1,000~{\rm km~s}^{-1}$, {\it assuming Keplerian motion}. Furthermore, deviations from a single Gaussian profile, such as exponential wings or Lorentzian-like shapes, indicate contributions from electron scattering and the presence of dense gas rather than purely virial kinematics \citep{Rusakov_2025,Chang_2026_Balmer,Scholtz2026arXiv260322277S}.

If the properties of LRDs are interpreted by analogy with stellar atmospheres, such as those of red-super giants and protostars, considering the role of magnetic fields is a natural extension. In such systems, magnetic fields are commonly generated through dynamo processes operating in convective layers, and similar conditions are expected in BHEs. This motivates us to investigate whether magnetized BHE models are consistent with current observations and how these fields affect their observational signatures. We focus on two key observables: the absence or weakness of X-ray emission and the formation of broadened hydrogen Balmer lines. The magnetic field of accreting objects can reduce the X-ray luminosity by decreasing the accretion shock temperature \citep[e.g.][]{Lamb1973ApJ,Elsner1977ApJ}. Conversely, a magnetized atmosphere may radiate strong X-rays, as commonly found in young low mass stars \citep{Gudel2004A&ARv}. Furthermore, if the BHEs are rotating, gas trapped in the corotating magnetosphere can produce Doppler broadening of emission lines, as suggested by studies of low mass stars \citep[e.g.][]{Cang2020A&A,Daley-Yates2024MNRAS}. Considering these effects, we aim to clarify the role of BHE magnetic fields.

The remainder of this paper is organized as follows. Section~\ref{sec:model_description} provides general descriptions of our model, summarizing the key structures and quantities. We also address the conditions for magnetosphere formation and the general structure of an accretion shock. Observational predictions regarding the formation of broad emission lines and the X-ray luminosity are presented in Section~\ref{sec:model_predictions}. In Section~\ref{sec:discussion}, we discuss limitations and implications of our model, emphasizing the importance of collaborative studies of stellar observations. Finally, Section~\ref{sec:summary} summarizes the key findings of this study.

\section{BH-Envelope-as-a-protostar Model} \label{sec:model_description}
\subsection{General Descriptions} \label{subsec:general}

\begin{figure}
\centering
\includegraphics[width=0.95\columnwidth]{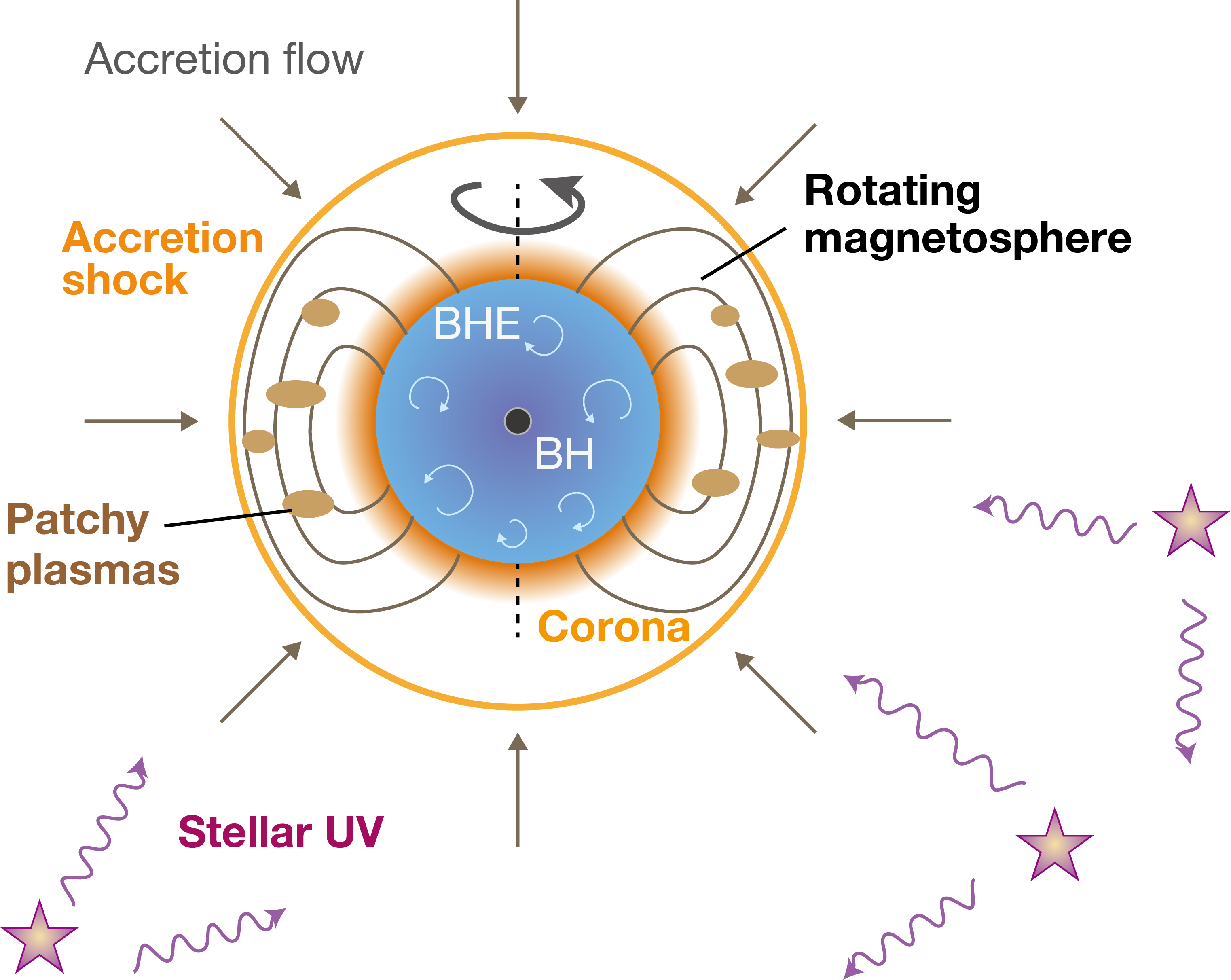}
\caption{Illustration of our BHE model. \label{fig:lrd-illust}}
\end{figure}

Figure~\ref{fig:lrd-illust} illustrates the overall structure of our LRD model. A massive black hole of mass $M_{\rm BH}$ is embedded within a BHE, which is analogous to a giant star or protostar near the Hayashi track on the Hertzsprung–Russell diagram \citep{Hayashi1961PASJ_a,Hayashi1961PASJ_b,Kido2025MNRAS,Begelman2026ApJ,H_Liu2025ApJ}. We assume that the BHE continues to grow via approximately spherical, free-fall accretion at a rate $\dot{M}_{\rm ISM}$, motivated by the expectation that LRDs host rapidly growing black holes. The BHE fully enshrouds the central black hole, thereby obscuring UV and X-ray radiation produced by AGN activity. This assumption differs from the scenario proposed by \citet{Naidu_2025} and \citet{Sneppen2026arXiv260118864S}, where AGN radiation can partially escape through the envelope. Given its proximity to the Hayashi track, the envelope is expected to be convective. Additionally, the envelope will possess finite angular momentum supplied by the accretion flow. Under these conditions, the combination of convection and rotation is likely to drive dynamo action, leading to the generation of large-scale, dipolar magnetic fields.

Magnetic fields affect the accretion structure by exerting the Lorentz force on the accreting flow, which forms a quasi-spherical accretion shock above the BHE photosphere. The post-shock layer is generally geometrically thin. Within the accretion shock, the plasma motion will be arrested by the magnetic fields. We define this region as {\it magnetosphere}. This magnetosphere rotates with the BHE and exhibits substructures, as described in Section~\ref{subsec:formulation}.

The convective and magnetized BHE is expected to form a hot, X-ray emitting corona, analogous to solar and stellar coronae. A fraction of the kinetic energy from convection is transported to the upper atmosphere via magnetic fields and subsequently dissipated magnetically in the tenuous region. If the atmosphere is dominated by magnetic energy, this dissipation can efficiently heat the tenuous gas.

\subsection{Model Formulation} \label{subsec:formulation}
This section describes four key assumptions and basic equations governing our BHE-as-a-protostar model. Given that our understanding of the internal structure of BHEs remains limited, we adopt a simplified framework where (i) a hydrostatic BHE is embedded within a dense, spherically symmetric free-fall accretion flow (Figure~\ref{fig:lrd-illust}). Under rapid mass accretion, the inflowing gas is expected to form an extended photosphere outside the hydrostatic envelope, analogous to protostellar accretion solutions of \citet{Stahler1986ApJ}. For simplicity, we assume that (ii) the photospheric radius and temperature are approximately equal to those at the envelope equilibrium surface, i.e., $r_{\rm ph}\simeq r_{\rm env}$ and $T_{\rm ph}\simeq T_{\rm surf}$. We further assume that (iii) the BHE mass is comparable to or smaller than the BH mass, allowing us to neglect the self-gravity of the envelope. Under these assumptions, we solve for the structure, spanning from the spherical free-fall accretion flow to the interior of the hydrostatic BHE. First, we determine the location and physical quantities such as the temperature of the photosphere in the accretion flow under the boundary conditions that the optical depth equals unity and the photospheric luminosity equals the Eddington luminosity (assumption~iv). Using these photospheric values as outer boundary conditions, we solve the internal structure of the BHE down to the vicinity of the central black hole, seeking solutions consistent with assumption~(iii). 
We then estimate the magnetic field strength from the physical conditions at the BHE surface, although the dynamical effects of magnetic fields on the envelope structure itself are neglected in the present study.

In the following, we adopt the convention of $A_X=A/10^X$ in cgs units unless otherwise noted, where $A$ denotes a physical quantity and $X$ is the exponent.

\subsubsection{Photosphere}

To effectively survey a broad range of $M_{\rm BH}$, we introduce the normalized variables $m_{\rm BH,7}=M_{\rm BH}/10^7M_{\odot}$ and $\dot{m}_{\rm ISM,1}=\dot{M}_{\rm ISM}/10\dot{M}_{\rm Edd}$. The quantity $\dot{M}_{\rm Edd}$ is defined as the Eddington accretion rate for a radiative efficiency $\eta(=0.1)$:
\begin{align}
    \dot{M}_{\rm Edd}=\frac{1}{\eta}\frac{4\pi GM_{\rm BH}}{c \kappa_{\rm T}} \approx 0.25~M_\odot~{\rm yr^{-1}}m_{\rm BH,7},
\end{align}
where $G$, $c$, and $\kappa_{\rm T}$ are the gravitational constant, the speed of light, and the Thomson opacity, respectively. For the primordial gas, we adopt $\kappa_T=0.35~{\rm cm^2/g}$. We note that in our normalization unit, $\dot{m}_{\rm ISM,1}=0.1$ corresponds to the Eddington accretion rate. The Eddington luminosity is calculated as 
\begin{align}
    L_{\rm Edd}\approx 1.4\times 10^{45}~m_{\rm BH,7}~{\rm erg~s^{-1}}.
\end{align}

\begin{figure}
\includegraphics[width=1.0\columnwidth]{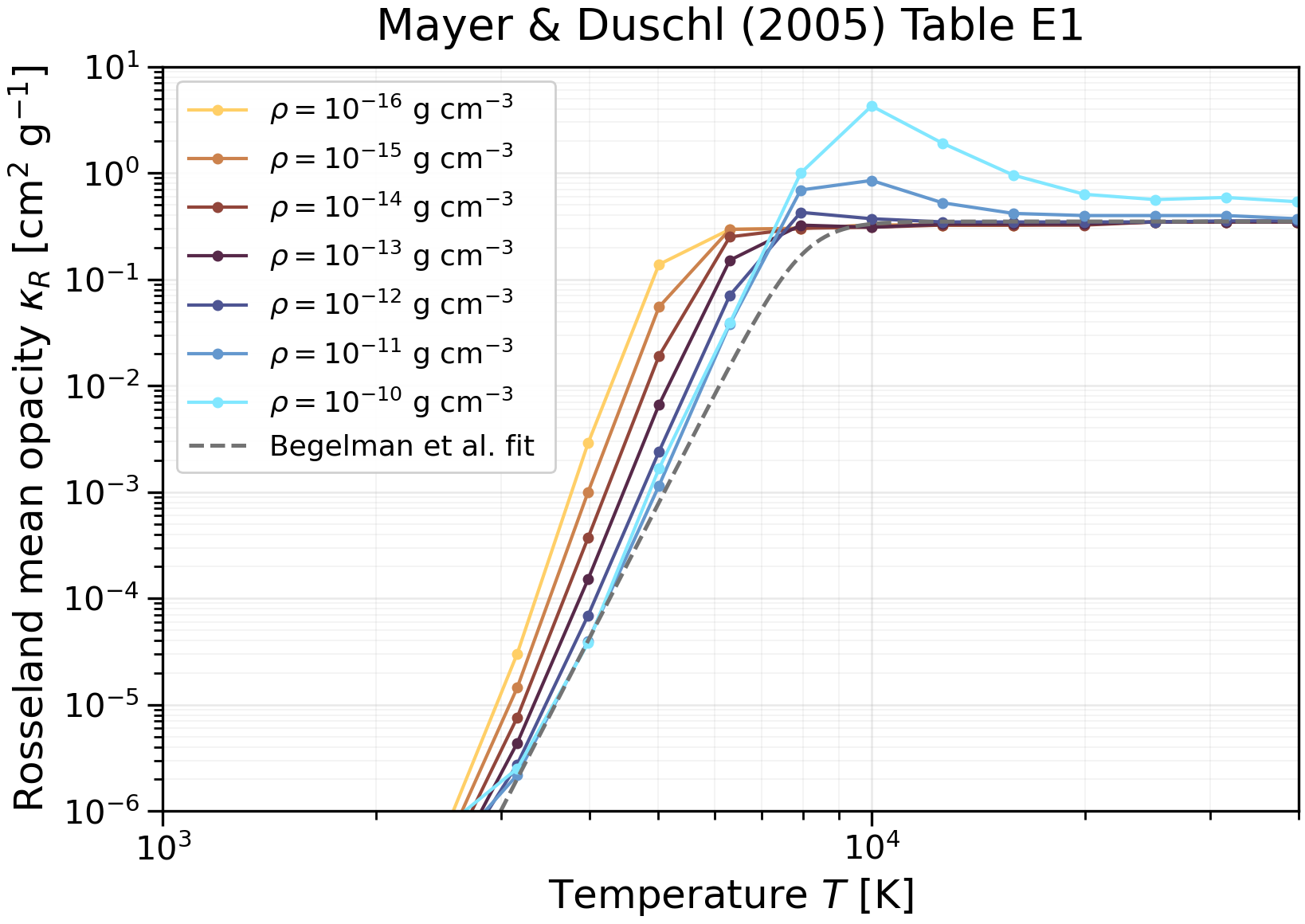}
\caption{The Rosseland mean opacity, taken from \citet{Mayer2005MNRAS}, is presented. The fitting function by \citet{Begelman2008MNRAS} is also included as a gray dashed line (Equation~(\ref{eq:opacity})). \label{fig:opacity}}
\end{figure}

The Rosseland mean opacity $\kappa$ is a key quantity for determining photospheric properties. Based on the opacity data for pristine gas \citep{Mayer2005MNRAS}, \citet{Begelman2008MNRAS} presented an approximate fitting formula that neglects density dependence:
\begin{align}
    \kappa(T)= \frac{\kappa_{\rm T}}{1+(T/T_0)^{-s}}, \label{eq:opacity}
\end{align}
where $T_0=8000~{\rm K}$, $s=13$, and $\kappa_{\rm T}=0.35~{\rm cm^2~g^{-1}}$. Its strong temperature dependence in the low temperature regime ($T<T_0$) is due to H$^{-}$ opacity. \citet{Kido2025MNRAS} adopted this function to derive the scaling relations of $r_{\rm ph}$ and $T_{\rm ph}$. However, the original opacity data in \citet{Mayer2005MNRAS} {\it does} show a non-negligible density dependence, as shown in Figure~\ref{fig:opacity}. The fitting function of Equation~(\ref{eq:opacity}) is also plotted for comparison. We note that the fitting function significantly underestimates the opacity data for low densities ($\rho \lesssim 10^{-12}~{\rm g~cm^{-3}}$) in $5000~{\rm K}\lesssim T\lesssim 7000~{\rm K}$. To mitigate the discrepancy between the fit and the tabulated data, we utilize the \citet{Mayer2005MNRAS} table data in this study. Therefore, the opacity $\kappa$ depends on both the density and temperature, $\kappa(\rho,T)$.

\begin{figure*}
\includegraphics[width=1.05\columnwidth]{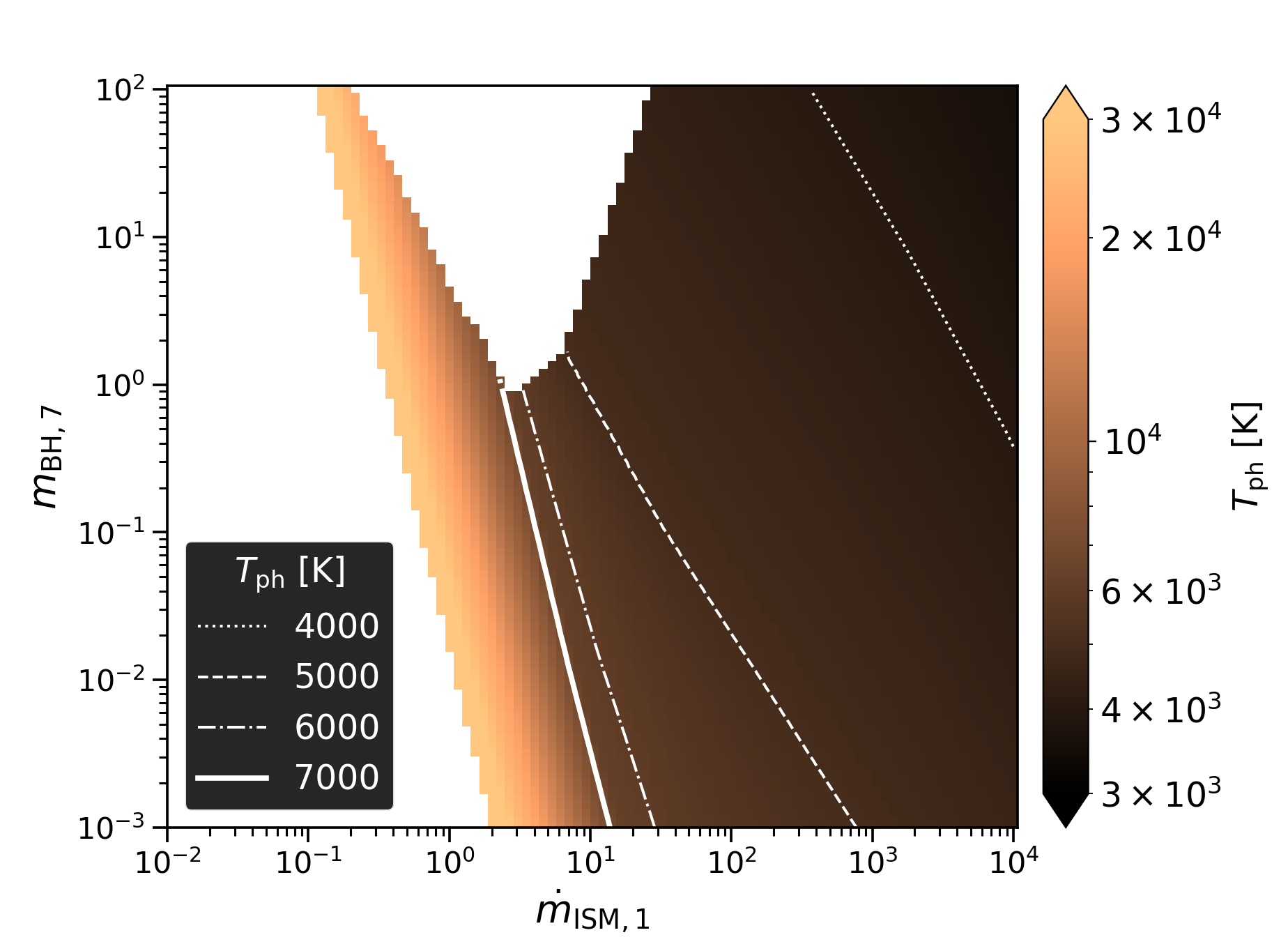}
\includegraphics[width=1.05\columnwidth]{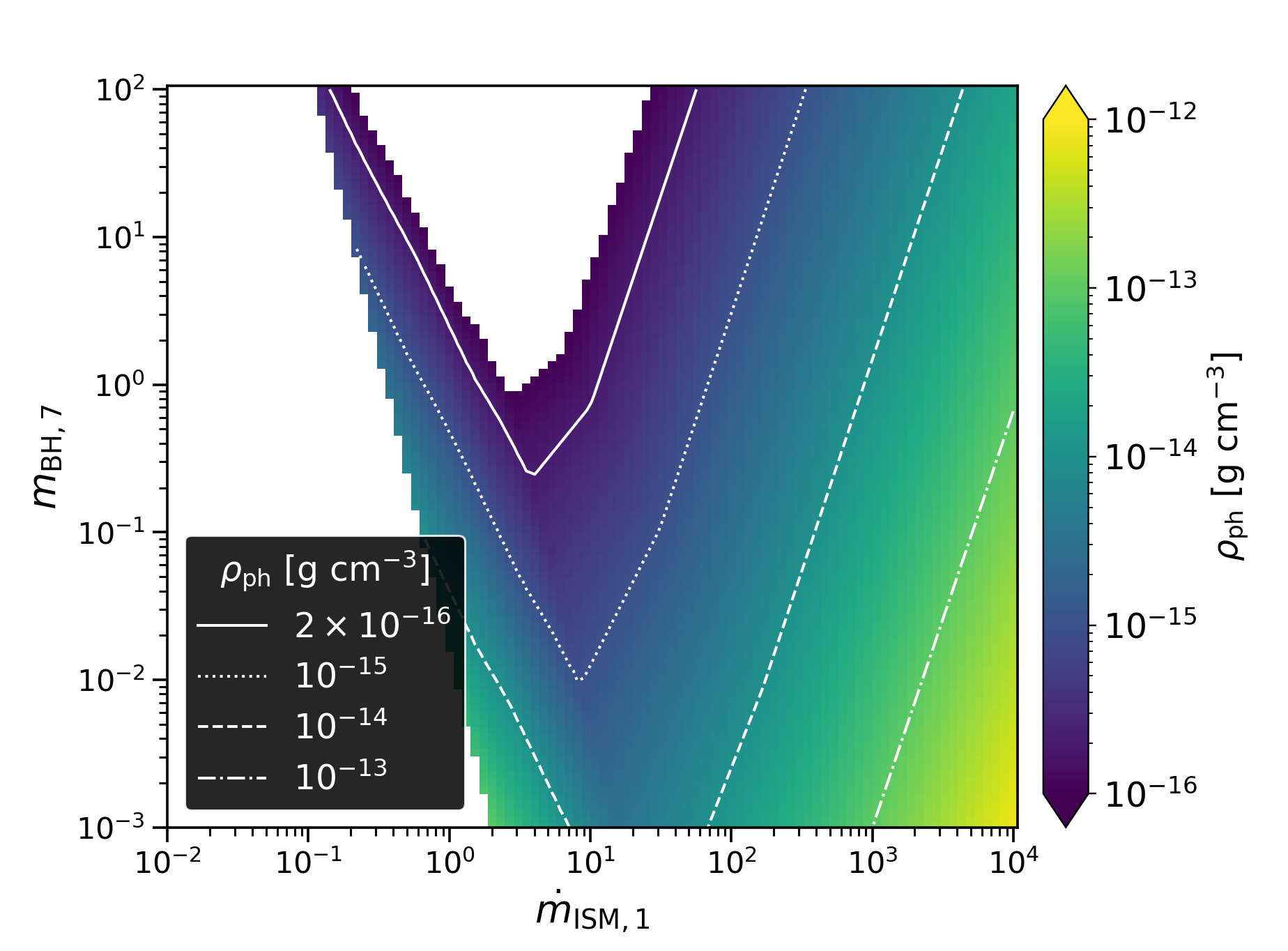}
\caption{Photosphere survey results derived using the \citet{Mayer2005MNRAS} opacity table. The left panel displays the photospheric temperature $T_{\rm ph}$ corresponding to the solutions for various sets of parameters defined by $(m_{\rm BH,7},\dot{m}_{\rm ISM,1})$. Regions marked blank indicate that no solutions exist under the given assumptions and opacity table. The right panel presents the map of the photospheric density $\rho_{\rm ph}$. \label{fig:solution_space}}
\end{figure*}

We numerically determine the photospheric solutions, $r_{\rm ph}$ and $T_{\rm ph}$, under assumptions (i) and (iv). Specifically, the photospheric conditions are given by \citep{Stahler1986ApJ,Kido2025MNRAS}:
\begin{align}
    \tau_{\rm ph}&=\rho_{\rm ph} \kappa(\rho_{\rm ph},T_{\rm ph})r_{\rm ph}=1,\\
    \rho_{\rm ph} &=\frac{\dot{M}_{\rm ISM}}{4\pi r_{\rm ph}^2 v_{\rm ff}(r_{\rm ph})},\\
    L_{\rm Edd} &=4\pi r_{\rm ph}^2 \sigma_{\rm SB} T_{\rm ph}^4,
\end{align}
where $v_{\rm ff}(r)=\sqrt{2GM_{\rm BH}/r}$ is the free fall velocity at a radius of $r$. The results are presented in Figure~\ref{fig:solution_space}. We find photospheric solutions within the colored region, while no solutions exist in the blank areas. Failure to find solutions occurs when the solver exits the defined parameter space (primarily in regions like $\rho_{\rm ph}\lesssim 10^{-16}~{\rm g~cm}^{-3}$) or fails to converge on a root under the given assumptions. Based on observed LRD spectra, photospheric solutions with temperatures of $T_{\rm ph}\lesssim 7000$~K are considered plausible \citep{deGraaff_2025b, Umeda2026ApJ}. Therefore, to restrict the analysis to the temperature range and avoid areas lacking solutions, we focus our study on the parameter space defined by $10\lesssim \dot{m}_{\rm ISM,1}\lesssim 10^4$ and $10^{-2}\lesssim m_{\rm BH,7}\lesssim 1$.

We note that $r_{\rm ph}$ and $T_{\rm ph}$ are functions of $M_{\rm BH}$ and $\dot{M}_{\rm ISM}$ under the model assumptions. Within the limited parameter space mentioned above, we derive the following fitting functions:
\begin{align}
    r_{\rm ph}&\approx 4.6\times 10^{16}~{\rm cm}~m_{\rm BH,7}^{a_1}\dot{m}_{\rm ISM,1}^{b_1},\label{eq:r_ph}\\
    T_{\rm ph}&\approx 5.5\times 10^3~{\rm K}~m_{\rm BH,7}^{a_2}\dot{m}_{\rm ISM,1}^{b_2},\label{eq:T_ph}
\end{align}
where $a_1=0.552$, $b_1=0.0780$, $a_2=-0.0258$, and $b_2=-0.0390$. We confirmed that the photospheric luminosity $L_{\rm ph}$, derived from the fitting functions, is nearly equal to $L_{\rm Edd}$, which is consistent with assumption~(iv). The photospheric temperature $T_{\rm ph}$ is approximately 5000--6000~K and exhibits a weak dependence on both $m_{\rm BH,7}$ and $\dot{m}_{\rm ISM,1}$. This value is generally consistent with LRD spectra. These solutions, characterized by such low photospheric temperatures, result from high accretion rates ($\dot{m}_{\rm ISM,1}\gtrsim 1$--10), which is consistent with the expectation that BHEs are rapidly growing in mass.

Equations~(\ref{eq:r_ph}) and (\ref{eq:T_ph}) differ from those shown in \citet{Kido2025MNRAS} because we directly adopt the density-dependent opacity table provided by \citet{Mayer2005MNRAS}. 
While the power-law relations presented in Equations~(\ref{eq:r_ph}) and (\ref{eq:T_ph}) are useful, they are only valid within the limited parameter space because the opacity scaling changes drastically around $T=5000$~K (Figure~\ref{fig:opacity}). Given this limitation, we retain $r_{\rm ph}$ explicitly in the following physical relations and adopt a fiducial value of $r_{\rm ph}=10^{17}~{\rm cm}$.

\subsubsection{Hydrostatic Black Hole Envelope}
Using the photospheric solutions as outer boundary conditions, we numerically determine hydrostatic-equilibrium solutions for embedded BHEs that satisfy assumptions (ii) and (iii). The effects of feedback from the central engine are ignored in this model \citep[see also][for the arguments about the feedback effect]{Kido2025MNRAS}. 

In this framework, the hydrostatic equilibrium structure is determined by two parameters: the energy injection from an accreting black hole with luminosity, $\beta \equiv 1-L/L_{\rm Edd}$, and the efficiency of convective energy transport $\beta_{\rm c}$ (see more details in the Appendix of \citealt{Kido2025MNRAS})\footnote{When the temperature gradient is steeper than the adiabatic value and the region becomes convectively unstable, the energy flux is approximated by $F_{\rm conv}=\beta_{\rm c}pc_{\rm s}$, where $p$ is the total pressure and $c_{\rm s}$ is the isothermal sound speed.}. 
To find BHE solutions under the black hole gravitational potential, we adopt $(\beta,\beta_{\rm c})=(10^{-4},3\times 10^{-3})$ as a fiducial set of parameters while exploring the ranges of $10^{-5}\le \beta \le 10^{-3}$ and $10^{-3}\le \beta_{\rm c}\le 10^{-1}$ to examine the existence of solutions. We confirmed that the gas pressure at the BHE surface, which is a key quantity for determining the BHE field strength (see Section~\ref{sec:mag}), is insensitive ($\lesssim 0.01\%$) to the choice of $\beta$ and $\beta_{\rm c}$ within the range explored here.
We note that for $M_{\rm BH}=10^7M_\odot$, no solution satisfying assumption (iii) was found for $\dot{m}_{\rm ISM,1}\gtrsim 10^3$. 
The solutions commonly show a density inversion near the BHE surface, as often found in one-dimensional models for very luminous stars \citep[e.g.][]{Joss1973ApJ}, but we neglect its potential impact on the interior structure for simplicity.

\subsubsection{Magnetosphere}\label{sec:mag}

When BHEs are magnetized, their magnetospheres will affect accretion dynamics, as discussed in many accreting objects \citep[e.g.][]{Elsner1977ApJ,Romanova2015SSRv}. To investigate these magnetic effects, we estimate the dipole field strength $B_{\rm surf}$ at the BHE surface using the thermal equipartition field $B_{\rm eq}$, which is defined as the maximum field strength that can be supported by the pressure at the BHE surface $p_{\rm surf}$ \citep[e.g.][]{Parker1978ApJ,Spruit1976SoPh}:
\begin{align}
    B_{\rm surf}=B_{\rm eq}=\sqrt{8\pi p_{\rm surf}}.
\end{align}
The total pressure at the BHE surface is the sum of the gas and radiation pressures, $p_{\rm gas}$ and $p_{\rm rad}$. This can be expressed as: 
\begin{align}
    p_{\rm surf}=p_{\rm gas}+p_{\rm rad}\approx \frac{2}{3}\frac{g(r_{\rm ph})}{\kappa(\rho_{\rm surf},T_{\rm surf})},
\end{align}
where $g(r)$ denotes the gravitational acceleration at a radius of $r$, and radiation pressure is ignored due to its negligible contribution at the surface. The surface density and temperature are denoted by $\rho_{\rm surf}$ and $T_{\rm surf}$, respectively, and we set $T_{\rm surf}=T_{\rm ph}$. Furthermore, we assume that the surface gravitational acceleration $g_{\rm ph}=g(r_{\rm ph})$ at the photosphere is primarily determined by the black hole mass ($g_{\rm ph}=GM_{\rm BH}/r_{\rm ph}^2$).

The assumption of thermal equipartition ($B_{\rm surf}=B_{\rm eq}$) is motivated by solar and stellar observations. In low-mass young stars (classical T Tauri stars, CTTSs), which are magnetized convective stars near the Hayashi limit, observations have often found field strengths similar to or larger than the thermal equipartition field by a factor of 2–3 \citep{Safier1999ApJ, Johns-Krull2007ApJ}. For giants, we examined Zeeman observations of \citet{Auriere2015A&A} and compared the observed field strength with $B_{\rm eq}$. As a result, most magnetically active giants in the sample show $B_{\rm surf}\sim B_{\rm eq}$, which generally supports our assumption. For more details on giants, see Appendix~\ref{app:bfield_giant}. Future magnetic field observations of BHEs will examine the validity of this assumption.

The magnetospheric radius $r_{\rm mag}$ quantifies the size of the magnetosphere and is determined by balancing the ram pressure exerted by the accretion flow against the magnetic pressure of the BHE dipole field:
\begin{align}
    \frac{B(r_{\rm mag})^2}{8\pi} \approx\rho_{\rm acc}(r_{\rm mag}) v_{\rm ff}(r_{\rm mag})^2,
\end{align}
where $\rho_{\rm acc}(r_{\rm mag})$ and $v_{\rm ff}(r_{\rm mag})$ denote the mass density of the accretion flow and the free-fall velocity, respectively, at a radius of $r_{\rm mag}$. Details regarding the magnetic geometry are ignored in this calculation. Using the relations given by $B(r)=B_{\rm surf}(r_{\rm ph}/r)^3$ and $\dot{M}_{\rm ISM}=4\pi r^2 \rho_{\rm acc}v_{\rm ff}$, we obtain the following expression \citep[e.g.][]{Lamb1973ApJ}:
\begin{align}
    r_{\rm mag}=k\left(\frac{B_{\rm surf}^4 r_{\rm ph}^{12}}{2GM_{\rm BH}\dot{M}_{\rm ISM}^2}\right)^{1/7},
\end{align}
where $k\sim \mathcal{O}(1)$ is a constant depending on the magnetic geometry. If $r_{\rm mag}>r_{\rm ph}$, the magnetosphere develops outside the BHE, and the accretion shock forms just outside the magnetosphere (see panel~(a) of Figure~\ref{fig:lrd-illust}). Conversely, in cases of $r_{\rm mag}\le r_{\rm ph}$, the magnetosphere is crushed by the accretion flows, causing the accretion shock to form at the photosphere. In this scenario, the shock radius $r_{\rm sh}$ will be ${\rm max}(r_{\rm mag},r_{\rm ph})$.

The passage through the accretion shock creates a high-temperature plasma. The temperature just behind the shock, $T_{\rm ps}$, is given by
\begin{align}
    T_{\rm ps} &= \frac{3}{16}\frac{\mu m_{\rm H}}{k_{\rm B}}v_{\rm ff}'^2. \label{eq:T-post-shock}
\end{align}
$\mu$ denotes the mean molecular weight, which is approximately 0.5 for fully-ionized primordial gas. We adopt $v_{\rm ff}'=(4/3)v_{\rm ff}$ as the accretion velocity, considering the rising motion of the shock above the static BHE photosphere \citep{Sacco2010A&A}. Due to its high temperature (typically $>10^7~{\rm K}$), the shocked gas can emit X-rays. We will investigate the resulting X-ray luminosity from the post-shock region in Section~\ref{subsec:X-ray-shock}.

The shocked gas cools and falls onto the BHE. This infall motion will be regulated by the rotating magnetic fields. Within the magnetosphere, accreting matter will experience a centrifugal force. A key quantity characterizing the significance of this force is the corotation radius $r_{\rm cor}$, defined as
\begin{align}
    r_{\rm cor}=\left( \frac{G M_{\rm BH}}{\Omega_{\rm ph}^2}\right)^{1/3}.\label{eq:r_cor}
\end{align}
The centrifugal force can dominate over the gravitational force when
\begin{align}
    r_{\rm cor}< r_{\rm mag},\label{ineq:rcor_rmag}
\end{align}
otherwise the centrifugal force is weaker. Previous studies of accreting stars \citep[e.g.][]{Romanova2015SSRv} refer to these two regimes as {\it the propeller and non-propeller regimes}.

Previous magnetohydrodynamic (MHD) simulations of stars suggest that the magnetosphere contains multiple patchy plasmas (see Figure~\ref{fig:lrd-illust}). Three-dimensional (3D) MHD simulations of accreting stars commonly demonstrate that accreting flows penetrate the magnetosphere, forming patchy accretion structures due to Rayleigh-Taylor-type instabilities \citep[e.g.][]{Kulkarni2008MNRAS, Takasao2022ApJ, Takasao2025ApJ_a}. When a star rotates rapidly (i.e., in the propeller regime), it produces outflows that may prevent the formation of such patchy accretion flows. However, gas outflowing from the stellar surface can accumulate around the equatorial plane inside the magnetosphere, forming dense clumps or disk-like structures \citep{ud-Doula2008MNRAS,Daley-Yates2024MNRAS}. Considering these findings from stellar models, we therefore expect the formation of patchy plasmas in the rotating magnetosphere over a wide range of BHE spin rates. We assume that these patchy plasmas are irradiated by starburst UV and consequently produce Balmer emission lines.

\subsubsection{Summary of formulae}
We summarize key quantities below. The field strength $B_{\rm surf}$ and magnetospheric radius $r_{\rm mag}$ depend on the opacity at the BHE surface. To illustrate this dependency, we introduce a normalized opacity $\tilde{\kappa}=\kappa/(5\times10^{-3}~{\rm cm^2~g^{-1}})$. This normalization unit approximately corresponds to the value in the \citet{Mayer2005MNRAS} table, specifically at $\rho=10^{-10}~{\rm g~cm^{-3}}$ and $T=5500$~{\rm K}.
\begin{align}
    g_{\rm ph}\approx 1.3\times 10^{-1}~{\rm cm~s^{-2}}~m_{\rm BH,7}r_{\rm ph,17}^{-2},
\end{align}
\begin{align}
    v_{\rm ff,ph}\approx1.6\times 10^3~{\rm km~s^{-1}}~m_{\rm BH,7}^{1/2}r_{\rm ph,17}^{-1/2},
\end{align}
\begin{align}
    \rho_{\rm ph} & \approx \frac{\dot{M}_{\rm ISM}}{4\pi r_{\rm ph}^2v_{\rm ff}(r_{\rm ph})}\nonumber\\
    &\approx 7.8\times 10^{-18}~{\rm g~cm^{-3}}~m_{\rm BH,7}^{-1/2}\dot{m}_{\rm ISM,1}r_{\rm ph,17}^{-3/2},
\end{align}
\begin{align}
        B_{\rm surf}&\approx \sqrt{\frac{16\pi}{3}\frac{g_{\rm ph}}{\kappa}}\nonumber \\
    &\approx 2.1\times 10~{\rm G}~\tilde{\kappa}^{-1/2}m_{\rm BH,7}^{1/2} r_{\rm ph,17}^{-1}
\end{align}
\begin{align}
    r_{\rm mag} & \approx 4.4\times 10^{17}~{\rm cm}~\tilde{\kappa}^{-2/7}m_{\rm BH,7}^{1/7}\dot{m}_{\rm ISM,1}^{-2/7}r_{\rm ph,17}^{8/7}\label{eq:r_mag}
\end{align}
\begin{align}
    \frac{r_{\rm mag}}{r_{\rm ph}}&\approx
    4.4~\tilde{\kappa}^{-2/7}m_{\rm BH,7}^{1/7}\dot{m}_{\rm ISM,1}^{-2/7}r_{\rm ph,17}^{1/7}\label{eq:rmag_rph}
\end{align}
These variables are fundamentally functions of only $M_{\rm BH}$ and $\dot{M}_{\rm ISM}$ (or $m_{\rm BH,7}$ and $\dot{m}_{\rm ISM,1}$), as $\kappa$ and $r_{\rm ph}$ are functions of these quantities in our framework.

\begin{figure}
\includegraphics[width=1\columnwidth]{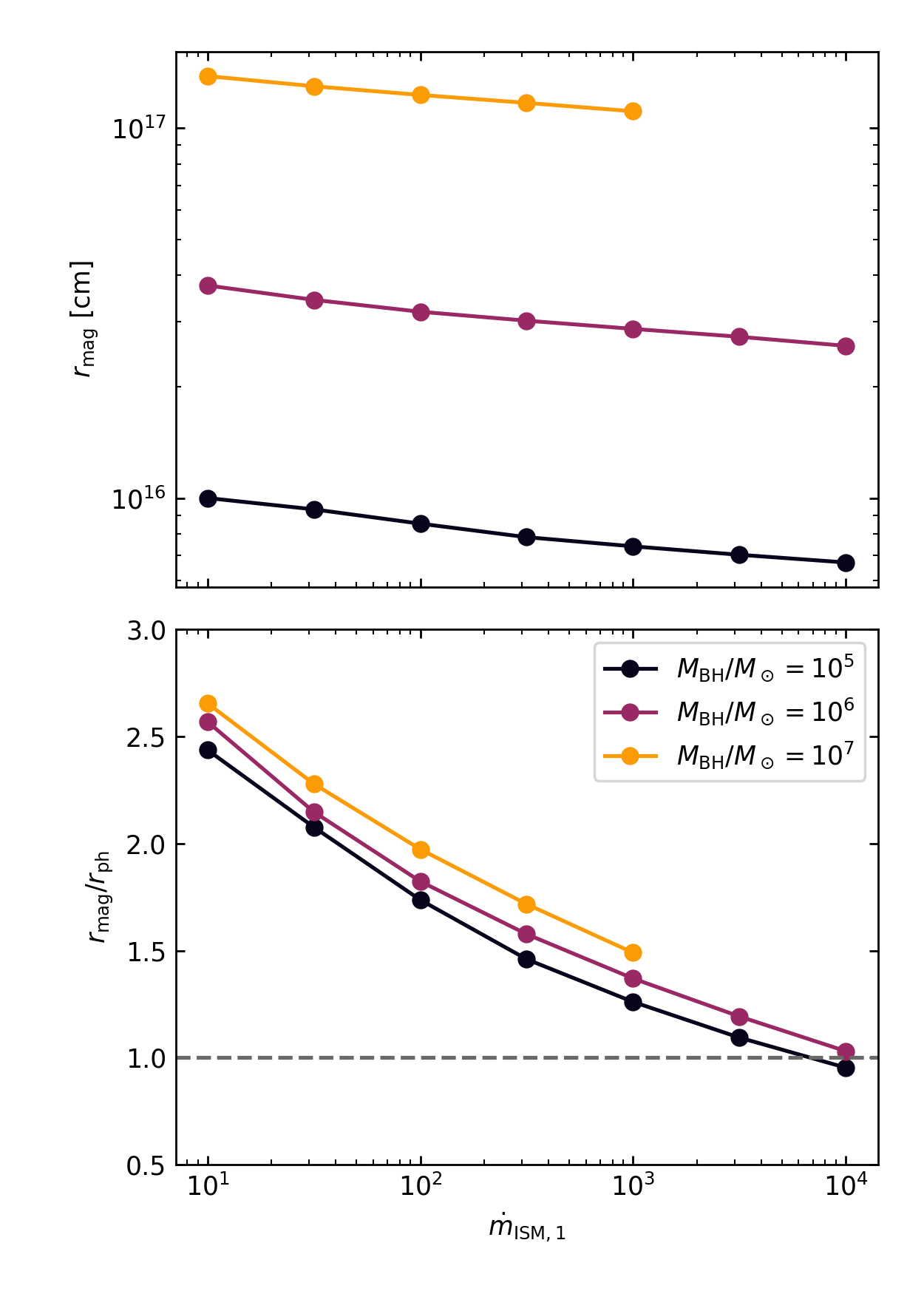}
\caption{The top and bottom panels illustrate the magnetospheric radius $r_{\rm mag}$ and the ratio $r_{\rm mag}/r_{\rm ph}$, respectively, as functions of the normalized accretion rate $\dot{m}_{\rm ISM,1}$. Results for different $M_{\rm BH}$ are presented. In the bottom panel, the dashed horizontal line marks a ratio of unity; above this threshold, the magnetosphere develops around the BHE. \label{fig:rmag_rph}}
\end{figure}

Figure~\ref{fig:rmag_rph} illustrates the dependence of $r_{\rm mag}$ (top panel) and the ratio of $r_{\rm mag}/r_{\rm ph}$ (bottom panel) on the normalized accretion rate $\dot{m}_{\rm ISM,1}$. Magnetosphere formation occurs when this ratio exceeds unity. The magnetospheric radius $r_{\rm mag}$ monotonically increases with $m_{\rm BH,7}$, while the ratio $r_{\rm mag}/r_{\rm ph}$ exhibits a weaker dependence on it. As $\dot{m}_{\rm ISM,1}$ increases, the magnetospheric size shrinks due to strong compression from the accretion flow. Also, the ratio approaches unity because the photospheric radius also expands, as shown in Equation~(\ref{eq:r_ph}). Consequently, the BHE model develops a magnetosphere across a wide range of parameters explored in this work.

\subsection{Key Questions of This Study}
Gas-enshrouded AGN scenarios, such as the BHE model, explain the low or absent X-ray emission observed in LRDs. However, if the BHE is indeed actively accreting, it may produce detectable X-ray emissions. 
When the accretion shock forms at the photospheric radius, the shock temperature will be
\begin{align}
T_{\rm ps, ph} &\equiv T_{\rm ps}(r_{\rm ph})\nonumber\\
& \approx 5.4\times 10^7~{\rm K}~m_{\rm BH,7}r_{\rm ph,17}^{-1}\label{eq:Tps_ph}
\end{align}
indicating that the shocked gas can emit X-rays due to its high temperature. 
The accretion luminosity $L_{\rm acc}$ is
\begin{align}
    L_{\rm acc}&\approx \frac{GM_{\rm BH}\dot{M}_{\rm ISM}}{r_{\rm ph}} \nonumber\\
    & \approx 2.1\times 10^{42}~{\rm erg~s^{-1}}~m_{\rm BH,7}\dot{m}_{\rm ISM,1}r_{\rm ph, 17}.
\end{align}
Given that the current observational upper limit for the X-ray luminosity is $\sim 10^{41}~{\rm erg~s^{-1}}$ \citep{Lin_2026}, either the efficiency of X-ray radiation must be very low, a substantial fraction of the X-ray emission must be absorbed before reaching observers, or both. We investigate the X-ray emissions from the post-shock region in Section~\ref{subsec:X-ray-shock}.

Although the surface conditions of BHEs remain highly uncertain, magnetized BHEs may produce hot, X-ray emitting atmospheres, analogous to the solar corona (Figure~\ref{fig:lrd-illust}). This modeled corona must also be consistent with observational constraints on the X-ray luminosity. We aim to clarify the allowed parameter space for the magnetized corona in Section~\ref{subsec:corona}.

Broad emission line profiles, such as the hydrogen H$\alpha$ line, are generally complex but often exhibit a Gaussian component with a FWHM of a few $100~{\rm km~s^{-1}}$ or larger, indicating contributions from Doppler broadening \citep[e.g.,][]{Rusakov_2025}. If BHEs obscure the central black hole, the standard AGN interpretation of line broadening based on virial motions of clouds around the black hole \citep{Reines2013ApJ} may not be applicable. Instead, we propose that gas clumps within the rotating magnetosphere play a key role in shaping these broad emission lines, and we explore this possibility in Section~\ref{subsec:rotation}.

\section{Model Predictions} \label{sec:model_predictions}

\subsection{SED of Accretion Shock} \label{subsec:accretion_shock}

When the magnetosphere develops, the accretion shock forms just outside it, where the gravitational potential is shallower than at the BHE photosphere. Consequently, the accretion speed and shock temperature at the magnetospheric radius, $v_{\rm ff,mag}$ and $T_{\rm ps,mag}$, are smaller than the corresponding quantities at the photospheric radius, $v_{\rm ff,ph}$ and $T_{\rm ps,ph}$, respectively. Under the condition of $T_{\rm ph}<T_0$, we evaluate:
\begin{align}
v_{\rm ff,mag} & \approx 7.8\times 10^{2}~{\rm km~s^{-1}}~\tilde{\kappa}^{1/7}m_{\rm BH,7}^{3/7}\dot{m}_{\rm ISM,1}^{1/7}r_{\rm ph,17}^{-4/7},\\
T_{\rm ps, mag} &\approx 1.2\times 10^7~{\rm K}~\tilde{\kappa}^{2/7}m_{\rm BH,7}^{6/7}\dot{m}_{\rm ISM,1}^{2/7}r_{\rm ph,17}^{-8/7}. \label{eq:T_ps_mag}
\end{align} 
Therefore, the BHE magnetic fields affects the SED of accretion-origin emissions.

To quantitatively investigate the SED and resulting X-ray luminosity from the post-shock region, we construct an analytical model for the shocked gas. A detailed description of this model is provided in Takasao et al. (in prep.). Here, we outline a general scheme for modeling the post-shock region: We assume that the post-shock region within the range of $T\ge 10^{4}~{\rm K}$ is optically thin. Calculating the radiatively cooling gas requires an optically thin radiative cooling function corresponding to a given elemental abundance. To obtain this function, we utilize the CHIANTI database \citep{DelZanna2021ApJ} and ChiantiPy (v0.15.2). For analytical calculations, we fit the resulting cooling function using a multi-power-law function across the temperature range of $T\ge 10^4~{\rm K}$. Subsequently, by applying the strong shock approximation and assuming constant gas pressure \citep[e.g.][]{Aoyama2018ApJ}, we analytically solve for the post-shock structure given an accretion speed and density. This solution yields the emission measure distribution (EMD), which is the emission measure as a function of temperature. Using the EMD, we compute the spectral luminosity as a function of wavelength, $L_\lambda~{\rm erg~s^{-1}~{\text{\AA}^{-1}}}$, via Chiantipy tools. We consider free-free, bound-free, and bound-bound emissions; for primordial gas composition, the free-free component dominates in the X-ray band. Finally, we obtain the total X-ray luminosity $L_{\rm X}$ within a given energy band by integrating $L_{\lambda}$. The detailed investigation of this X-ray luminosity will be presented in Section~\ref{subsec:X-ray-shock}.

\begin{figure*}
\centering
\includegraphics[width=1.6\columnwidth]{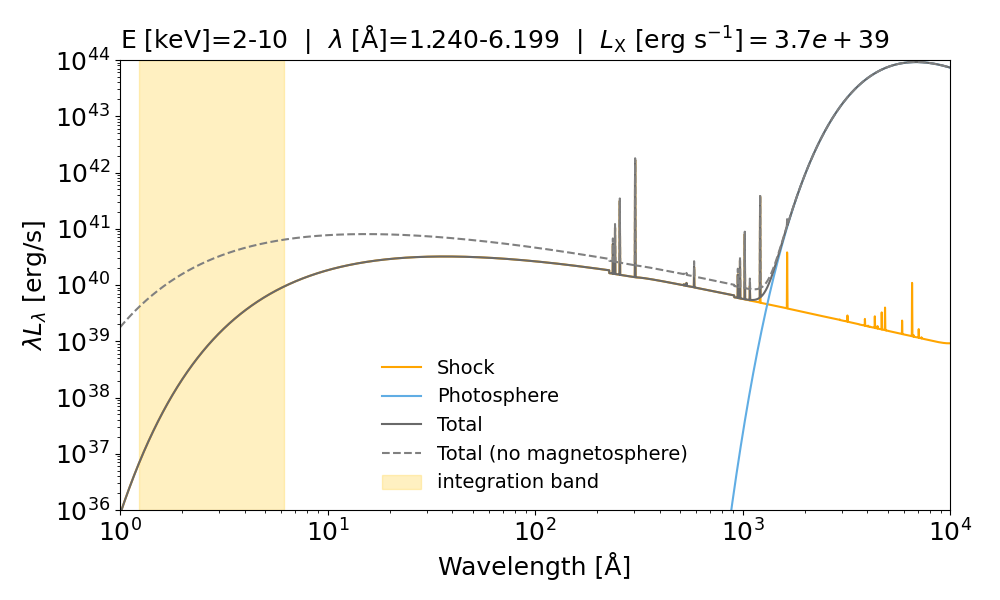}
\caption{Intrinsic spectral luminosity of our BHE model calculated for $M_{\rm BH}=10^6 M_\odot$ and $\dot{M}_{\rm ISM}=100\dot{M}_{\rm Edd}$ (or $m_{\rm BH,7}=0.1$ and $\dot{m}_{\rm ISM,1}=10$). The blue line represents the blackbody radiation from the photosphere, characterized by a temperature of $5249$~K. The orange line denotes the total emission from the post-shock region, while the dimgray line shows the sum of these two components. We assume a primordial gas composition (H and He only). The X-ray luminosity in the photon energy range of $2\text{-}10$~keV is displayed at the top panel. The corresponding wavelength range used for computing this X-ray luminosity is highlighted by the orange shaded region. Additionally, the total spectrum derived from the unmagnetized model is shown as a gray dashed curve. \label{fig:shock_spec_lum}}
\end{figure*}

\begin{figure*}
\centering
\includegraphics[width=1.6\columnwidth]{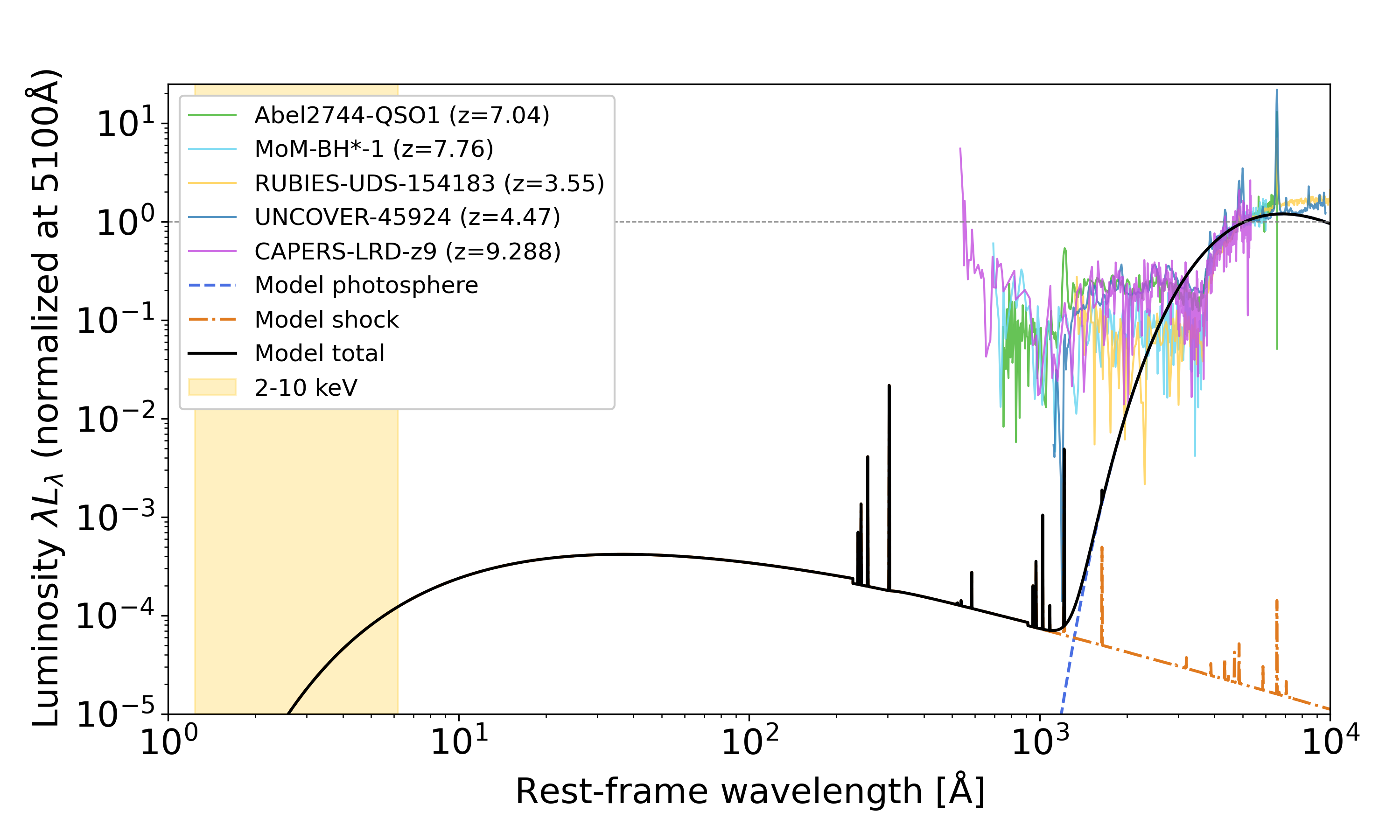}
\caption{Comparison of the intrinsic spectrum of our BHE model with observational data. Both the modeled and observed spectral luminosities ($\lambda L_{\lambda}$) are normalized at rest-frame 5100\AA. The modeled spectral luminosity is calculated using the same parameters as those presented in Figure~\ref{fig:shock_spec_lum} ($m_{\rm BH,7}=0.1$ and $\dot{m}_{\rm ISM,1}=10$). Overlaid observational data for LRDs include: A2744-QSO1 at $z = 7.04$ \citep{Furtak2024Natur}, MOM-BH$^*$-1 at $z=7.76$ \citep{Naidu_2025}, RUBIES-UDS-154183 at $z=3.55$ \citep{deGraaff_2025b}, UNCOVER-45924 at $z=4.47$ \citep{Labbe_2024b}, and CAPERS-LRD-z9 at $z=9.288$ \citep{Taylor_2025b}. These data are also used in \citet{Inayoshi_2025d}. \label{fig:sed_obs_model}}
\end{figure*}

Figure~\ref{fig:shock_spec_lum} illustrates the spectral luminosity calculated by our post-shock model. The figure displays the blackbody radiation from the BHE photosphere (blue), the total emission from the post-shock region (orange), and the sum of these two components (dimgray). We assume primordial elemental abundances, consisting only of H and He. The normalized BH mass and accretion rates are set to $m_{\rm BH,7}=0.1$ and $\dot{m}_{\rm ISM,1}=10$, respectively. The accretion shock forms at the magnetospheric radius $r_{\rm mag}\approx 2.6r_{\rm ph}$ (see Figure ~\ref{fig:rmag_rph}).
The spectral component on the longer wavelength is formed by photospheric radiation, while the shortward spectrum originates from the optically thin emission of the post-shock gas. The negative slope observed in this region directly reflects the thermal structure of the post-shock plasma. The shocked gas temperature is $T_{\rm ps, mag}\approx 1.4\times 10^{7}~{\rm K}$ immediately behind the shock, decreasing downstream due to radiative cooling. Since hotter plasmas occupy larger volumes, the gas with a temperature of $\sim T_{\rm ps,mag}$ acts as the primary coolant in the post-shock region. Consequently, the spectral luminosity at longer wavelengths is generally lower due to the smaller volume occupied by cooler plasma.

Moreover, to illustrate the role of magnetic fields in shaping the spectrum, we show the non-magnetized model spectrum using the gray dashed curve in Figure~\ref{fig:shock_spec_lum}. In this magnetosphere-free scenario, the accretion shock forms at the photospheric radius $r_{\rm ph} (\le r_{\rm mag})$ and leads to the production of hotter plasmas compared to the magnetized case. Therefore, the magnetized model is predicted to yield much weaker X-ray emission. We investigate the properties of accretion-origin X-rays in detail in Section~\ref{subsec:X-ray-shock}.

Figure~\ref{fig:sed_obs_model} compares the model spectrum with observed spectral luminosities for several LRDs. All spectral luminosities are normalized by their values at rest-frame 5100~\AA. Our BHE model exhibits a large UV-to-optical ratio of $\sim O(10^{-3})$, whereas many LRDs show the ratio of $\mathcal{O}(0.1)$ \citep[e.g.][]{Inayoshi_Ho_2025}. Consequently, the observed UV continuum in LRDs must arise from additional energy sources, such as starburst activity surrounding the BHE and/or leakage of UV photons from the BHE interior through the polar regions (\citealt{Asada_2026}; see also \citealt{Inayoshi_2025d,Lin_2026,W.Sun_2026}). Although the post-shock region forms hot emission lines, their intensities are considerably weaker than the observed UV continuum and therefore appear obscured.

\subsection{Interpretation of Observed Broad Line Regions} \label{subsec:rotation}

\begin{figure}
\centering
\includegraphics[width=0.95\columnwidth]{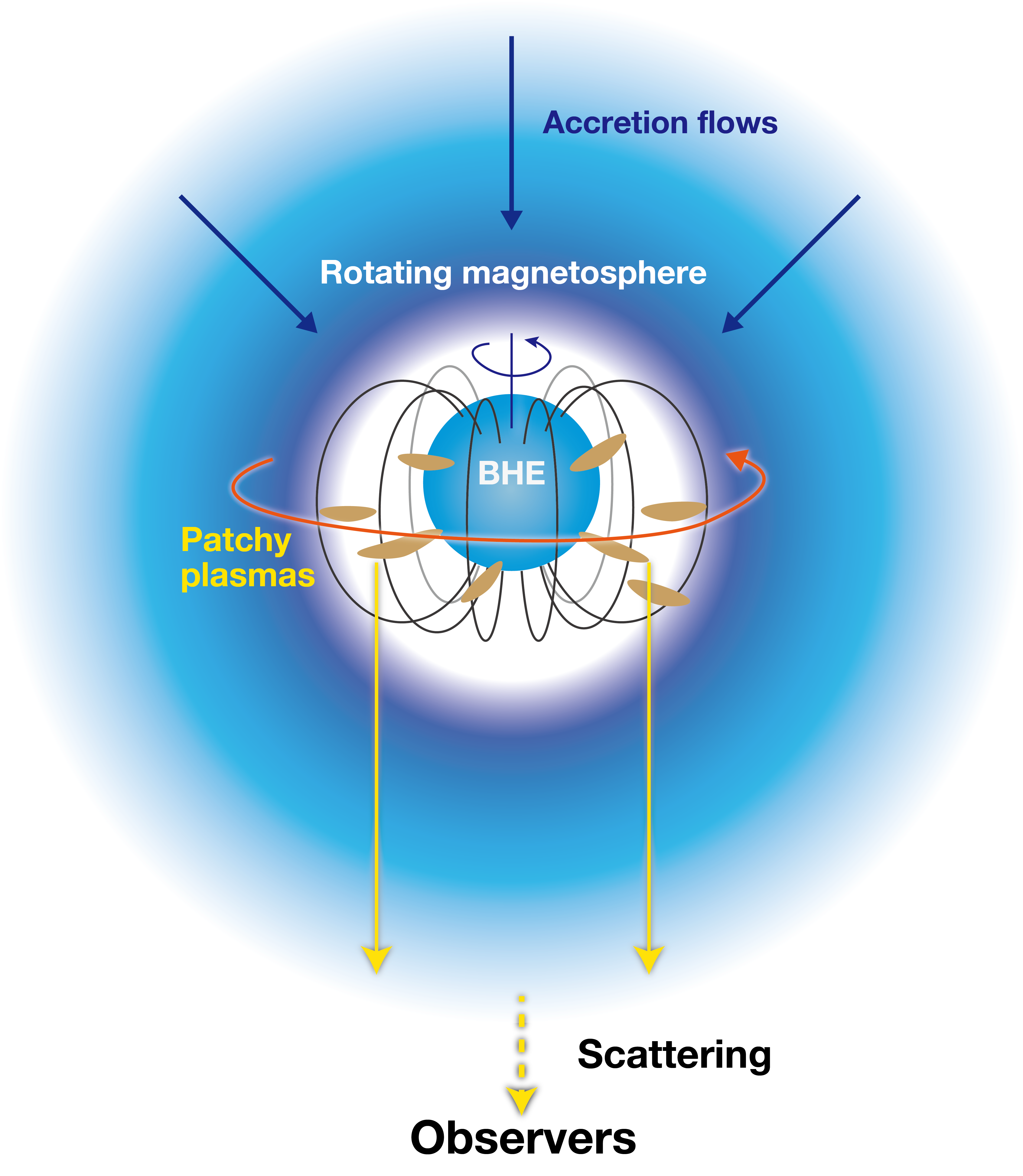}
\caption{Schematic illustration of broad emission line formation in our model. Yellow arrows highlight the emissions originating from plasma clouds corotating with the magnetosphere. \label{fig:lrd-illust-emission}}
\end{figure}

Figure~\ref{fig:lrd-illust-emission} illustrates a schematic view of line-emitting plasma clouds corotating with the magnetosphere (see also Section~\ref{subsec:general} for the conjecture about plasma clumps). The resulting emission lines are subject to Doppler shifting due to the corotation motion of these plasma clouds and potential scattering by electrons (originating from accreting flows or larger-scale gas ionized by starburst UV) before reaching observers. Consequently, the observed line profile is modulated by both the Doppler effect and electron scattering. Furthermore, depending on the column density of neutral hydrogen, Balmer absorption features will be superimposed upon the broad-line profiles.

\subsubsection{Line profiles}

\begin{figure*}
\centering
\includegraphics[width=0.95\columnwidth]{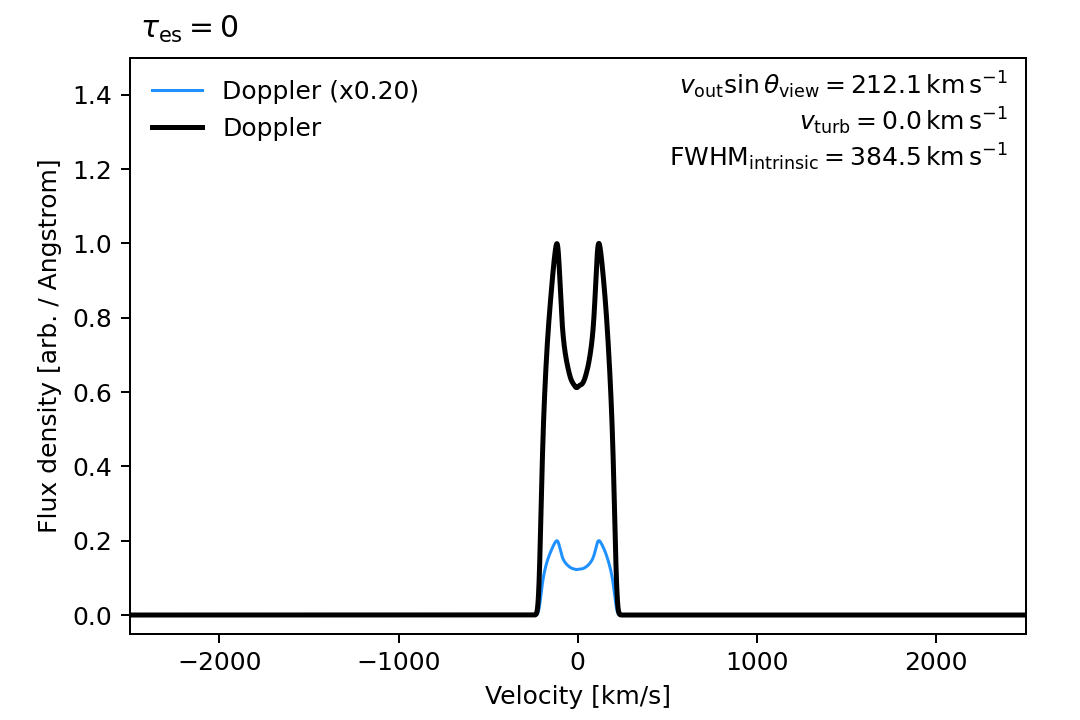}
\includegraphics[width=0.95\columnwidth]{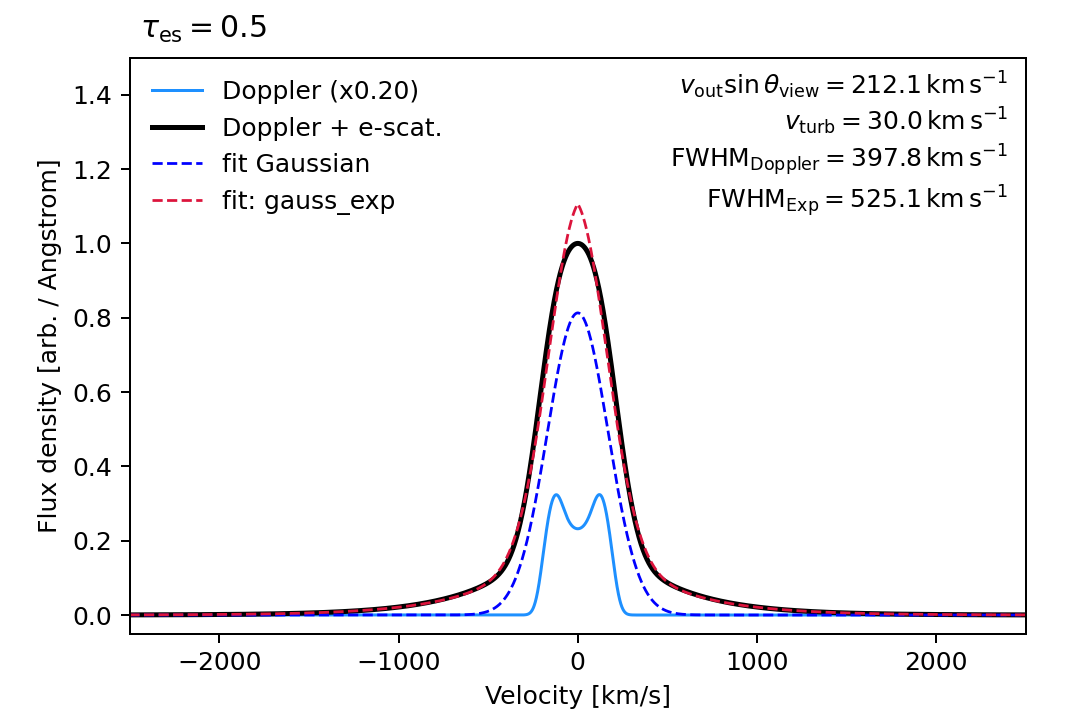}
\includegraphics[width=0.95\columnwidth]{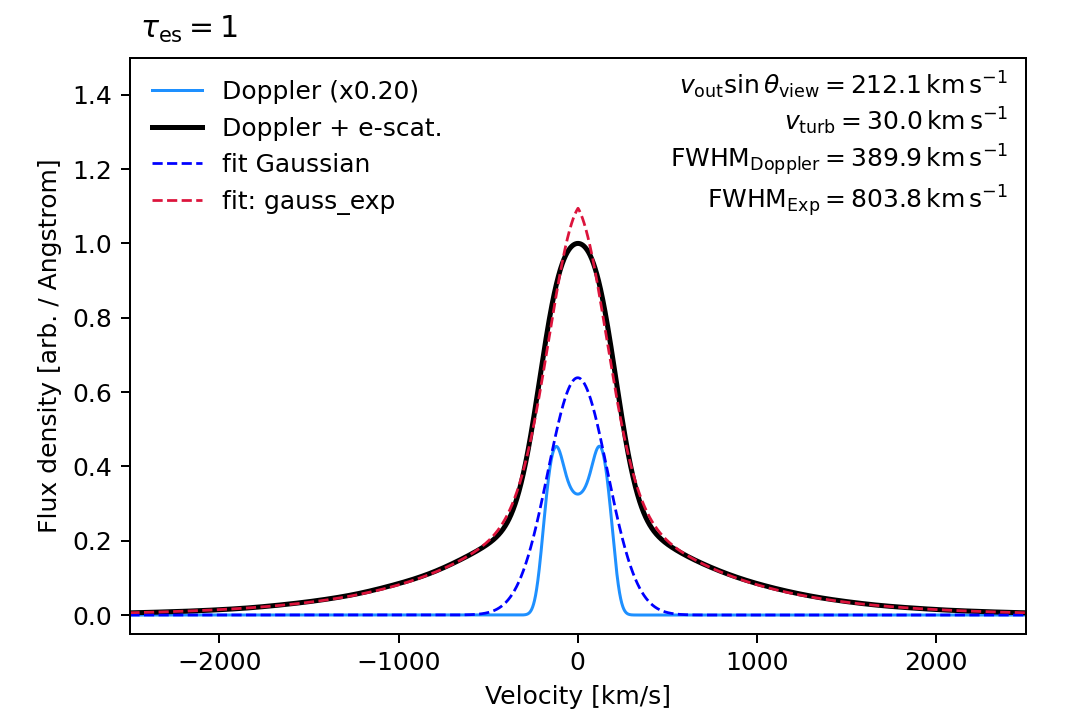}
\includegraphics[width=0.95\columnwidth]{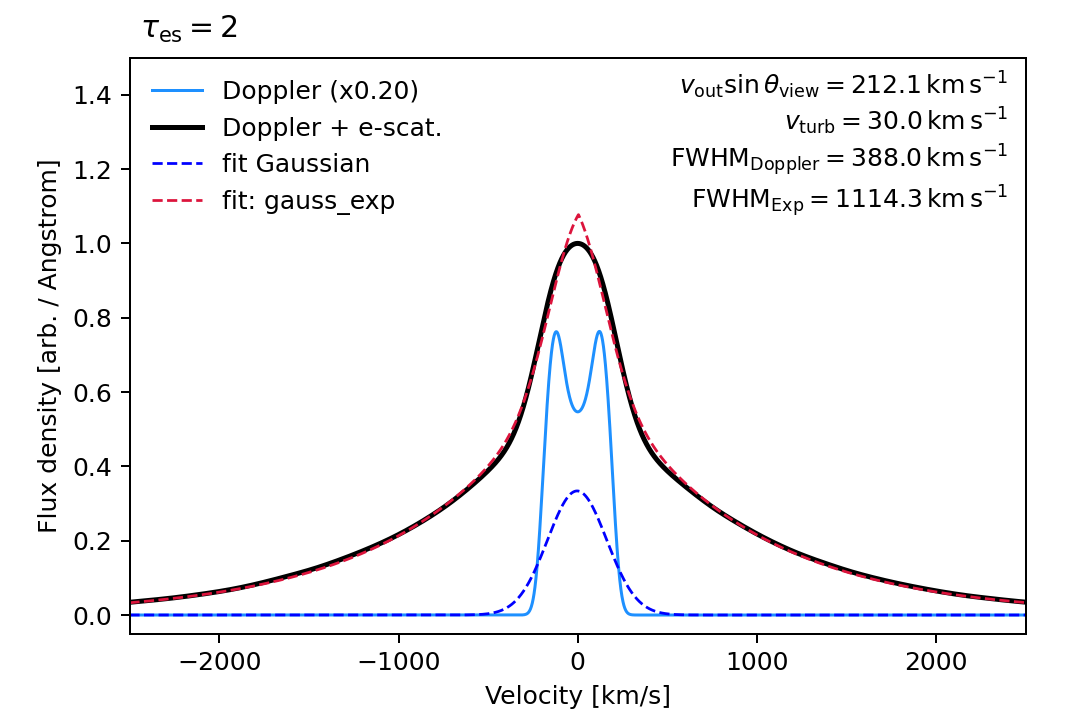}
\caption{Schematic examples of expected H$\alpha$ line shapes broadened by Doppler effects and electron scattering. The top-left panel presents the reference line profile, calculated assuming a rotation velocity at the outer disk edge of $v_{\rm out}=300~{\rm km~s^{-1}}$ and a viewing angle of $\theta_{\rm view}=45^\circ$ from the rotation axis, considering only the Doppler effect. In contrast, the panels on the top right, bottom left, and bottom right incorporate additional effects: electron scattering, turbulence, and spectral smoothing. For these examples, the electron scattering optical depth is varied as $\tau_{\rm es}=$0.5, 1, and 2. Furthermore, a smoothing width of $100~{\rm km~s^{-1}}$ is applied to mimic limited spectral resolution. Within these panels, the solid blue lines show profiles modified only by Doppler effects from rotation and turbulence (scaled by a factor of 0.2). The black solid lines represent the full spectral profiles, including Doppler effects, electron-scattering broadening, and smoothing. The red dashed lines denote the overall fitting function (a Gaussian plus an exponential tail), while the blue dashed lines indicate the fitted Gaussian components. Detailed model descriptions and parameters are provided in the main text. \label{fig:line-profile-demo}}
\end{figure*}

To illustrate schematic examples of line profiles broadened by Doppler effects and electron scattering, we perform line-shape calculations using a simplified model. To mimic emission from rotating gas clumps within the magnetosphere, we assume a {\it rigidly rotating disk} that produces an emission line. The ratio of inner to outer radii is treated as a model parameter that controls the velocity contrast within the magnetosphere, and the disk is assumed to have a symmetric emissivity profile. Furthermore, we include line broadening due to turbulent motions to account for disordered plasma cloud movements. The effect of electron scattering is characterized by the scatterer gas temperature and the scattering optical depth, $\tau_{\rm es}$. We fix the gas temperature at $10^{4}~{\rm K}$, corresponding to an electron thermal velocity of approximately $400~{\rm km~s^{-1}}$. Limited spectral resolution is accounted for by smoothing the resulting profile. Our calculations utilize several free parameters: the ratio of inner to outer radii, the viewing angle $\theta_{\rm view}$, the rotation velocity at the outer disk edge $v_{\rm out}$, the turbulent velocity $v_{\rm turb}$, the optical depth $\tau_{\rm es}$, and the smoothing width. Using a Monte Carlo method, we compute the H$\alpha$ line profile as observed from a specified viewing angle.

Figure~\ref{fig:line-profile-demo} presents three examples illustrating different $\tau_{\rm es}$. In all models, we set the ratio of the inner to outer radii to 0.5 and the viewing angle to $\theta_{\rm view}=45^\circ$ relative to the rotation axis, $v_{\rm out}=300~{\rm km~s^{-1}}$. The top left panel shows the line profile modified solely by the Doppler effect, which exhibits a double-peaked structure. The intrinsic FWHM is $\textrm{FWHM}_{\rm intrinsic}\approx 385~{\rm km~s^{-1}}$, approximately $2v_{\rm out \sin{\theta_{\rm view}}}\approx 424~{\rm km~s^{-1}}$. We find the relation $\textrm{FWHM}_{\rm intrinsic}\approx 2v_{\rm out \sin{\theta_{\rm view}}}$ over a wide range of $v_{\rm out}$ and $\theta_{\rm view}$. The other panels display how this reference line profile changes due to the effects of electron scattering, turbulence, and spectral smoothing. We set $v_{\rm turb}=0.1v_{\rm out}=30~{\rm km~s^{-1}}$. The smoothing width is $100~{\rm km~s^{-1}}$, which approximately corresponds to a JWST/NIRSpec high resolution mode ($R\simeq 2700$ at redshift $z=0$). When electron scattering and smoothing are considered, the double-peaked structure is smeared out. The resulting profile can be fitted by a combination of a Gaussian-like core and an exponential tail. As $\tau_{\rm es}$ increases, the line wing broadens. We note that the double-peaked structure remains visible for $\tau_{\rm es}\lesssim 2$ when $v_{\rm out}$ is larger than several 100~${\rm km~s^{-1}}$.

To demonstrate the impact on observational interpretation, we model the line profiles using a combination of a Gaussian function and an exponential tail, following \citet{Rusakov_2025}. Specifically, the profile is fitted with the following function:
\begin{align}
    f_{\rm fit}(v)=A_1\exp{\left(\frac{v^2}{2\sigma_{\rm D}^2}\right)} + A_2 \exp{\left(-\frac{|v|}{W}\right)},
\end{align}
where $A_1$ and $A_2$ denote the amplitudes of the Gaussian and exponential tails, respectively. When we define the velocity dispersion of the Gaussian component as $\textrm{FWHM}_{\rm Doppler}$ and the e-folding scale of the exponential tail as $\textrm{FWHM}_{\rm Exp}$ (Figure~\ref{fig:line-profile-demo}), then $\sigma_{\rm D}$ and $W$ are related to these quantities as follows: $\textrm{FWHM}_{\rm Doppler}=2\sqrt{2\ln{2}}\sigma_{\rm D}$ and $\textrm{FWHM}_{\rm Exp}$ as $\textrm{FWHM}_{\rm Exp}=2\sqrt{\ln{2}}W$. It is worth noting that some studies, such as \citet{Matthee2026arXiv260317667M}, adopt a different fitting approach by considering an additional narrow Gaussian component separately. The fitting results show that $\textrm{FWHM}_{\rm Doppler}\approx 390\textrm{--}400~{\rm km~s^{-1}}$ and $\textrm{FWHM}_{\rm Doppler}$ are nearly insensitive to the scattering optical depth $\tau_{\rm es}$. Therefore, we find $\textrm{FWHM}_{\rm Doppler}\approx \textrm{FWHM}_{\rm intrinsic}\approx 2v_{\rm out}\sin{\theta_{\rm view}}$. In contrast, the line tail scale $\textrm{FWHM}_{\rm Exp}$ increases from $\sim 530~{\rm km~s^{-1}}$ to $\sim 1100~{\rm km~s^{-1}}$ as $\tau_{\rm es}$ increases from 0.5 to 2. These examples suggest that fitting the spectrum with a Gaussian function alone is insufficient for estimating the true Doppler component.

\subsubsection{Methods for BH Mass Estimate}
Historically, $\textrm{FWHM}_{\rm Doppler}$ and the luminosity have been combined to estimate the black hole mass in AGNs, assuming that line broadening is primarily produced by the Doppler effect. We briefly review this method; AGN observations reveal an empirical size-luminosity relation for broad line regions (BLRs) \citep{Benzt2013ApJ}:
\begin{align}
    r_{\rm BLR}\approx 9.3\times 10^{16}~{\rm cm}~\left(\frac{L_{\rm 5100}}{10^{44}~{\rm erg~s^{-1}}}\right)^{0.542},
\end{align}
where $L_{\rm 5100}$ is defined as the monochromatic luminosity at $5100$~\AA. Furthermore, $L_{5100}$ is considered a reasonable measure of the luminosity of ionizing photons produced in AGNs. To estimate the black hole mass, the virial relation has been commonly employed \citep{Greene2005ApJ, Reines2013ApJ}, assuming that the broad-line emitting clouds are virialized at the BLR radius:
\begin{align}
    \frac{GM_{\rm BH,vir}}{r_{\rm BLR}}=f_{\rm vir}(\textrm{FWHM}_{\rm Doppler})^2,
\end{align}
where $f_{\rm vir}$ is a virial coefficient, and we adopt $f_{\rm vir}=1$ as a fiducial value. We distinguish the true black hole mass, denoted by $M_{\rm BH}$, from the black hole mass estimated using the virial relation, $M_{\rm BH,vir}$. By combining these two relations, we find the following:
\begin{align}
    \textrm{FWHM}_{\rm Doppler}&\approx 1.2\times 10^3~{\rm km~s^{-1}}\nonumber \\
    &\times \left(\frac{M_{\rm BH,vir}}{10^7 M_\odot}\right)^{1/2}\left(\frac{L_{\rm 5100}}{10^{44}~{\rm erg~s^{-1}}}\right)^{-0.271}. \label{eq:dv_fwhm_obs}
\end{align}
Therefore, assuming the virial relation, we can observationally estimate the black hole mass $M_{\rm BH}$ using $\textrm{FWHM}_{\rm Doppler}$ and $L_{\rm 5100}$. $\textrm{FWHM}_{\rm Doppler}$ is often estimated from H$\alpha$ line emission, and \citet{Greene2005ApJ} presents an empirical relation between the broad H$\alpha$ luminosity and $L_{5100}$:
\begin{align}
    L_{\rm H\alpha}=5.25\times 10^{42}~{\rm erg~s^{-1}}\left(\frac{L_{5100}}{10^{44}~{\rm erg~s^{-1}}}\right)^{1.157}.\label{eq:Lhalpha-L5100}
\end{align}

Our BHE-as-a-protostar model proposes an alternative interpretation to the standard understanding. Within this framework, the photospheric luminosity is written as
\begin{align}
    L_{\rm ph}&=4\pi r_{\rm ph}^2 \sigma_{\rm SB} T_{\rm ph}^4 = L_{\rm Edd}.\label{eq:Lph}
\end{align}
We note that $L_{\rm ph}$ is independent of the accretion rate $\dot{m}_{\rm ISM,1}$, as demonstrated by $L_{\rm ph}= L_{\rm Edd}\propto M_{\rm BH}$. Furthermore, line broadening originates not from virial motion of gas clouds but from rotation within the magnetosphere. Based on the idealized model in Section~\ref{subsec:rotation}, we predict $v_{\rm out}=r_{\rm mag}\Omega_{\rm ph}$ or

\begin{equation}
    \textrm{FWHM}_{\rm Doppler}\approx 2 r_{\rm mag}\Omega_{\rm ph}\sin{\theta_{\rm view}}
\end{equation}
We parametrize the spin rate using a nondimensional parameter, $f_{\rm rot}$, defined as
\begin{align}
    \Omega_{\rm ph}=f_{\rm rot}\Omega_{\rm K}(r_{\rm ph}),\label{eq:definition_f_rot}
\end{align}
and the value must be restricted to the range $0\le f_{\rm rot}\le 1$ to avoid the break-up of the BHE due to centrifugal force. Consequently, we obtain the following relation: 
\begin{align}
    \textrm{FWHM}_{\rm Doppler}&\approx 2f_{\rm rot}\frac{r_{\rm mag}}{r_{\rm ph}}v_{\rm K}(r_{\rm ph})\sin{\theta_{\rm view}},\nonumber \\
    &\approx 1.0\times 10^3~{\rm km~s^{-1}} \sin{\theta_{\rm view}} f_{\rm rot,-1}\nonumber \\
    & \times \tilde{\kappa}^{-2/7}m_{\rm BH,7}^{9/14}\dot{m}_{\rm ISM,1}^{-2/7}r_{\rm ph,17}^{-5/14}. \label{eq:FWHM_model}
\end{align}
$\textrm{FWHM}_{\rm Doppler}$ of this model exhibits a weak dependence on both $m_{\rm BH,7}$ and $\dot{m}_{\rm ISM,1}$, but depends on the viewing angle $\theta_{\rm view}$. This equation can predict the possible maximum value for $\textrm{FWHM}_{\rm Doppler}$. Furthermore, it can be used to verify whether the BHE is rotating at a spin rate smaller than the break-up value.

\begin{figure*}
    \centering
    \includegraphics[width=1.0\columnwidth]{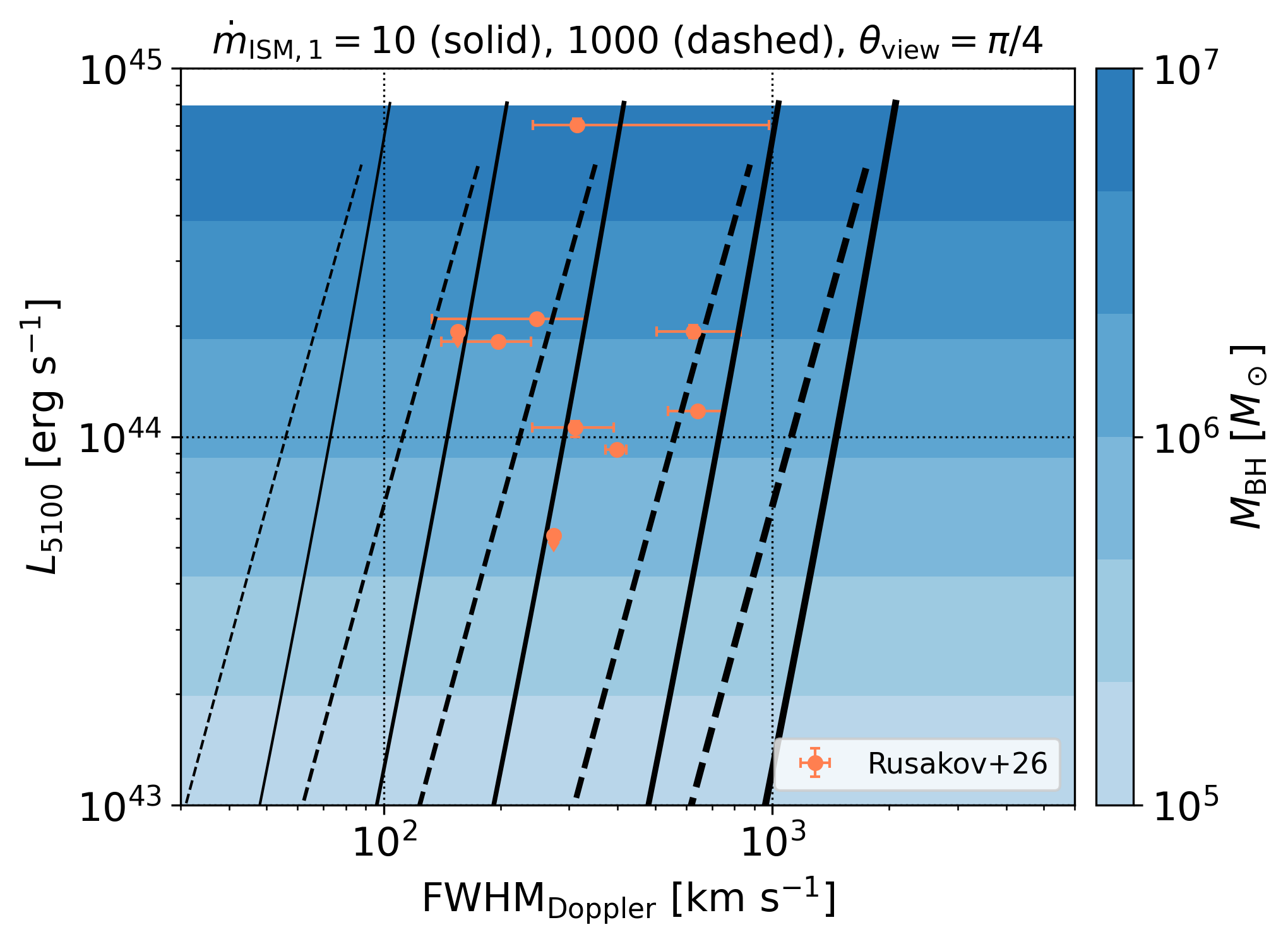}
    \includegraphics[width=1.0\columnwidth]{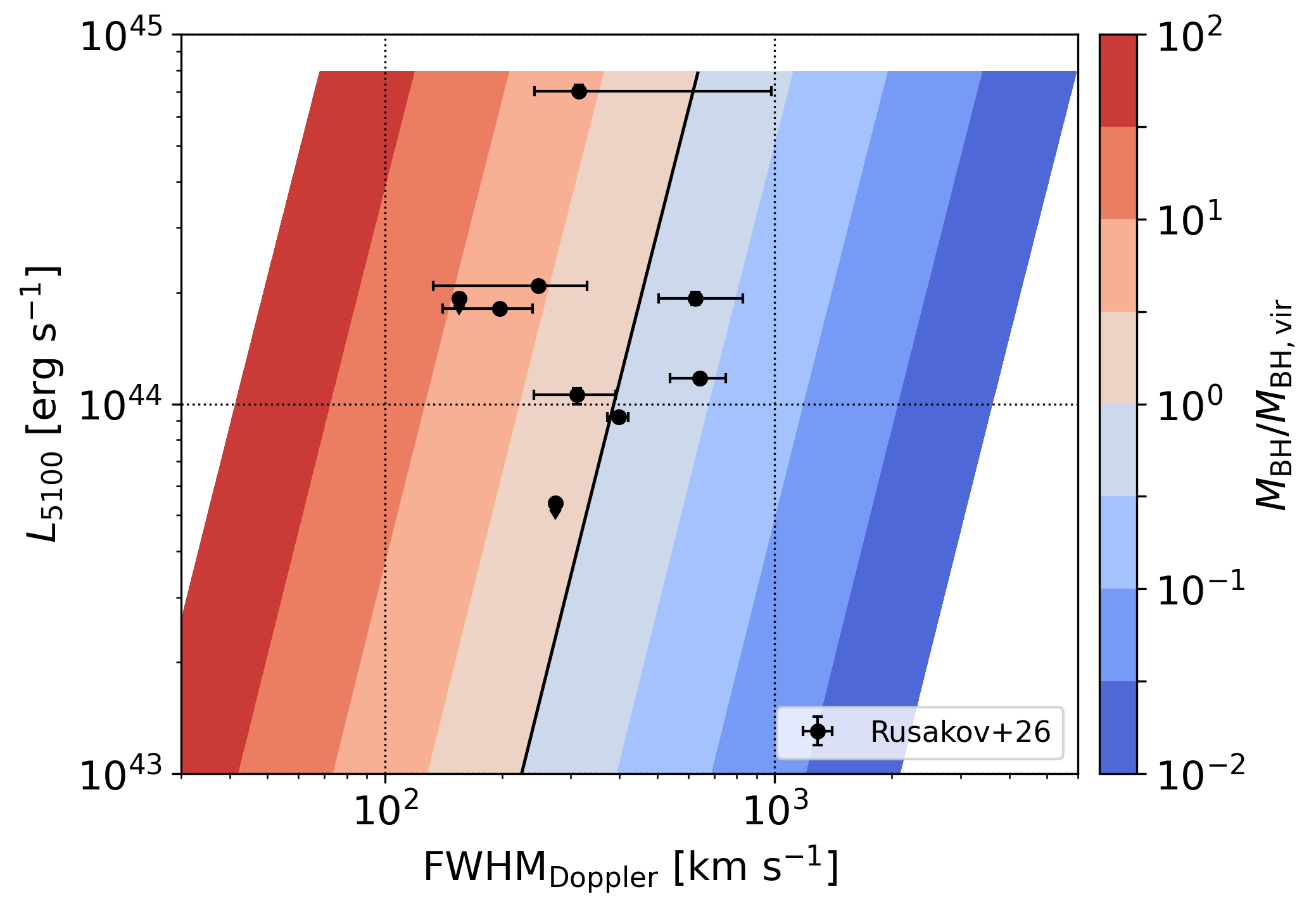}
    \caption{Left: Relation between $\textrm{FWHM}_{\rm Doppler}$ and $L_{5100}$ for $\theta_{\rm view}=\pi/4$. The color contours depict the black hole mass, $M_{\rm BH}$. Solid lines illustrate $\textrm{FWHM}_{\rm Doppler}$ for $\dot{m}_{\rm ISM,1}=10$, showing values for $f_{\rm rot}=0.03$, 0.1, 0.3, and 1 from thin to thick lines. The colored region is constrained by the explored range of the black hole mass ($10^5 \le M_{\rm BH}/M_{\rm odot}\le 10^7$). Dashed lines correspond to cases defined by $\dot{m}_{\rm ISM,1}=1000$. Right: Ratio of the theoretical black hole mass $M_{\rm BH}$ to that inferred using the virial relation $M_{\rm BH,vir}$. In both panels, points with error bars represent data sourced from Extended Data Table~3 in \citet{Rusakov_2025}. Note that the reported FWHM values include only the Doppler component, excluding contributions from electron scattering. Data for IDs B, D, and G are excluded due to complex line profiles, and data for ID~E is also excluded due to its high H$\alpha$ luminosity. Furthermore, conversion of $L_{\rm H\alpha}$ to $L_{5100}$ utilized Equation~(\ref{eq:Lhalpha-L5100}). \label{fig:FWHM-L5100}}
        \vspace{5mm}
\end{figure*}

The left panel of Figure~\ref{fig:FWHM-L5100} displays the relation between $\textrm{FWHM}_{\rm Doppler}$ and $L_{5100}$ derived from our model with $\theta_{\rm view}=\pi/4$. The color contours map the black hole mass $M_{\rm BH}$, which corresponds to $L_{5100}$ (Equation~\ref{eq:Lph}). Separately, the monochromatic luminosity $L_{5100}$ is calculated using the blackbody spectrum for $T_{\rm ph}$.
The solid curves depict $\textrm{FWHM}_{\rm Doppler}$ as defined in Equation~(\ref{eq:FWHM_model}) for the value of $\dot{m}_{\rm ISM,1}=10$, ranging from thin to thick lines corresponding to $f_{\rm rot}=0.03$, 0.1, 0.3, and 1. These curves are nearly vertical because $\textrm{FWHM}_{\rm Doppler}$ exhibits only a weak dependence on $m_{\rm BH,7}$, and consequently on $L_{5100}$. The range of $100~{\rm km~s^{-1}}\lesssim \textrm{FWHM}_{\rm Doppler}\lesssim 500~{\rm km~s^{-1}}$ implies $0.1\lesssim f_{\rm rot}\lesssim 0.3$. Our model is capable of accounting for $\textrm{FWHM}_{\rm Doppler}\approx 2000~{\rm km~s^{-1}}$ when $f_{\rm rot}\approx 1$. Furthermore, increasing the value of $\dot{m}_{\rm ISM,1}$ results in a slightly smaller $\textrm{FWHM}_{\rm Doppler}$ (shown by dashed lines).

The right panel of Figure~\ref{fig:FWHM-L5100} compares the black hole mass in our model with that estimated using the virial relation (Equation~\ref{eq:dv_fwhm_obs}). The plot indicates that when $L_{5100}$ is higher and $\textrm{FWHM}_{\rm Doppler}$ is smaller, the virial-relation-based method tends to underestimate $M_{\rm BH}$, and vice versa. Consequently, we caution against simply applying the virial-relation-based method for black hole mass estimation.

In both panels of Figure~\ref{fig:FWHM-L5100}, we incorporate LRD data of $\textrm{FWHM}_{\rm Doppler}$ which were carefully derived by \citet{Rusakov_2025}. While the observed FWHM is generally $>1000~{\rm km~s^{-1}}$, the study argues that line broadening is heavily affected by electron scattering \citep[see also][]{Begelman2026ApJ}. By subtracting this contribution, they derive a corrected value, $\textrm{FWHM}_{\rm Doppler}$. Furthermore, $L_{5100}$ is inferred from $L_{\rm H\alpha}$ using Equation~(\ref{eq:Lhalpha-L5100}). The data points are generally consistent with the model parameter ranges of $0.1\lesssim f_{\rm rot}\lesssim 0.3$ and $10^{6}M_\odot \lesssim M_{\rm BH}\lesssim 10^{7}M_\odot$ (left panel). 
Finally, the right panel suggests that applying the virial-relation-based method erroneously estimate the black hole mass of a fraction of the samples by a factor of 3-10.

\subsection{X-ray Luminosity}\label{subsec:X-ray}
One crucial observational property of LRDs is the weakness or absence of X-rays \citep[e.g.][]{Inayoshi_Ho_2025}. To test whether our model is consistent with observations, we investigate the intrinsic X-ray luminosity arising from an accreting and magnetized BHE. Following \citet{Lin_2026}, we adopt $10^{41}~{\rm erg~s^{-1}}$ as the observational upper limit for the X-ray luminosity in the energy range of $2 \text{-} 10$~keV.

We consider two primary sources for X-ray emission: (1) the post-shock region of the accretion shock, which we denote as the accretion-origin, and (2) the magnetically heated corona, designated as the magnetic-origin. These structures are illustrated in Figure~\ref{fig:lrd-illust}. We estimate the intrinsic X-ray luminosities from these two sources separately.

\subsubsection{Accretion-origin X-rays: Accretion Shocks} \label{subsec:X-ray-shock}

Using the spectral luminosity model of the post-shock region (Section~\ref{subsec:accretion_shock}), we estimate the X-ray luminosity produced by the accretion shock. Considering the case $(m_{\rm BH,7},\dot{m}_{\rm ISM,1})=(0.1,10)$ as an example (Figure~\ref{fig:shock_spec_lum}), we find that the shock temperature is $T_{\rm ps, mag}\approx 1.4\times 10^{7}~{\rm K}$. In this case, the X-rays with a photon energy of approximately $1~{\rm keV}$ (or $\sim 10$\AA) dominate the radiative output. The intrinsic X-ray luminosity in the energy range of $2\text{-}10$~keV is approximately $3.7\times 10^{39}~{\rm erg~s^{-1}}$ (see the colored region in Figure~\ref{fig:shock_spec_lum}).

\begin{figure}
\includegraphics[width=1.0\columnwidth]{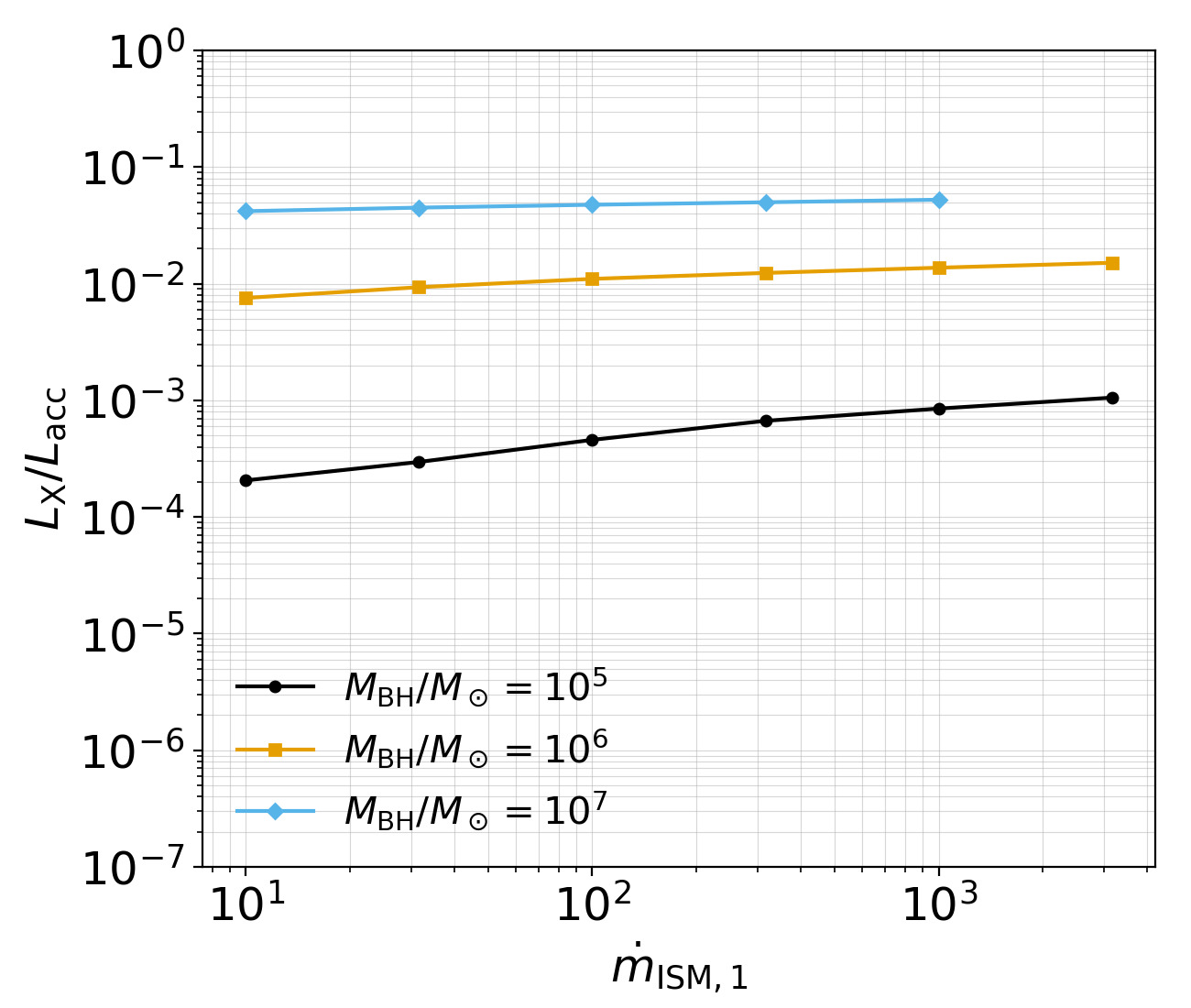}
\includegraphics[width=1.0\columnwidth]{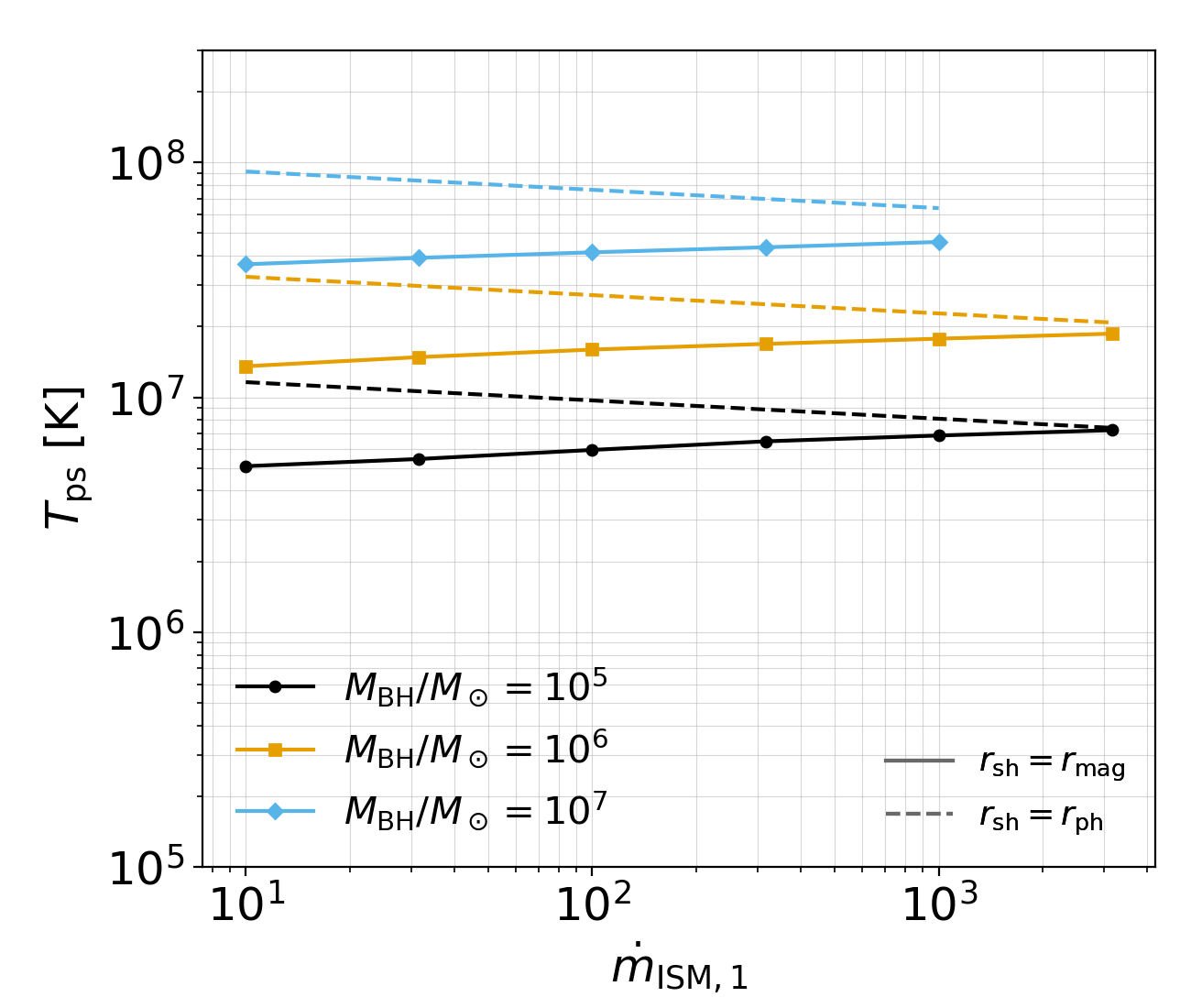}
\caption{Top panel: Radiative efficiency of X-rays in terms of the accretion luminosity, $L_{\rm X}/L_{\rm acc}$, as a function of $\dot{m}_{\rm ISM,1}$. The black, orange, and blue lines denote results for cases of $M_{\rm BH}=10^5,10^6, 10^7M_\odot$, respectively. Bottom panel: Shock temperature at the magnetospheric radius, $T_{\rm ps,mag}$, as a function of $\dot{m}_{\rm ISM,1}$. Solid and dashed curves show the results with and without magnetospheres.}
\label{fig:shock_Tps_Lx_over_Lacc}
\end{figure}

The dependence of the X-ray radiative efficiency on black hole mass and accretion rate is examined here. The top panel of Figure~\ref{fig:shock_Tps_Lx_over_Lacc} displays the ratio of intrinsic X-ray luminosity to the accretion luminosity, $L_{\rm X}/L_{\rm acc}$. It should be noted that X-ray attenuation is ignored in this calculation. This ratio increases sensitively with $m_{\rm BH,7}$, while exhibiting only a weak increasing trend with increasing $\dot{m}_{\rm ISM,1}$. Considering that the X-ray luminosity is sensitive to the amount of plasma with temperatures in the range of $1\text{--}10$~keV ($\sim 10^7\text{--}10^8~{\rm K}$), we also examine its dependence on the shock temperature $T_{\rm ps,mag}$. The bottom panel of Figure~\ref{fig:shock_Tps_Lx_over_Lacc} shows the dependency of $T_{\rm ps,mag}$. Specifically, the shock temperatures for the cases of $M_{\rm BH}/M_\odot=10^6$ and $10^7$ are above $10^7~{\rm K}$, whereas those for the cases of $M_{\rm BH}/M_\odot=10^5$ are below the temperature. This difference explains why the X-ray radiative efficiency for the cases of $M_{\rm BH}/M_\odot=10^5$ is significantly smaller than that found in the higher mass cases. Furthermore, the bottom panel compares shock temperatures with and without magnetospheres. The discrepancy between these two regimes is significant at lower accretion rates but decreases for higher rates as the magnetosphere shrinks (Figure~\ref{fig:rmag_rph}).

\begin{figure}
\includegraphics[width=1.0\columnwidth]{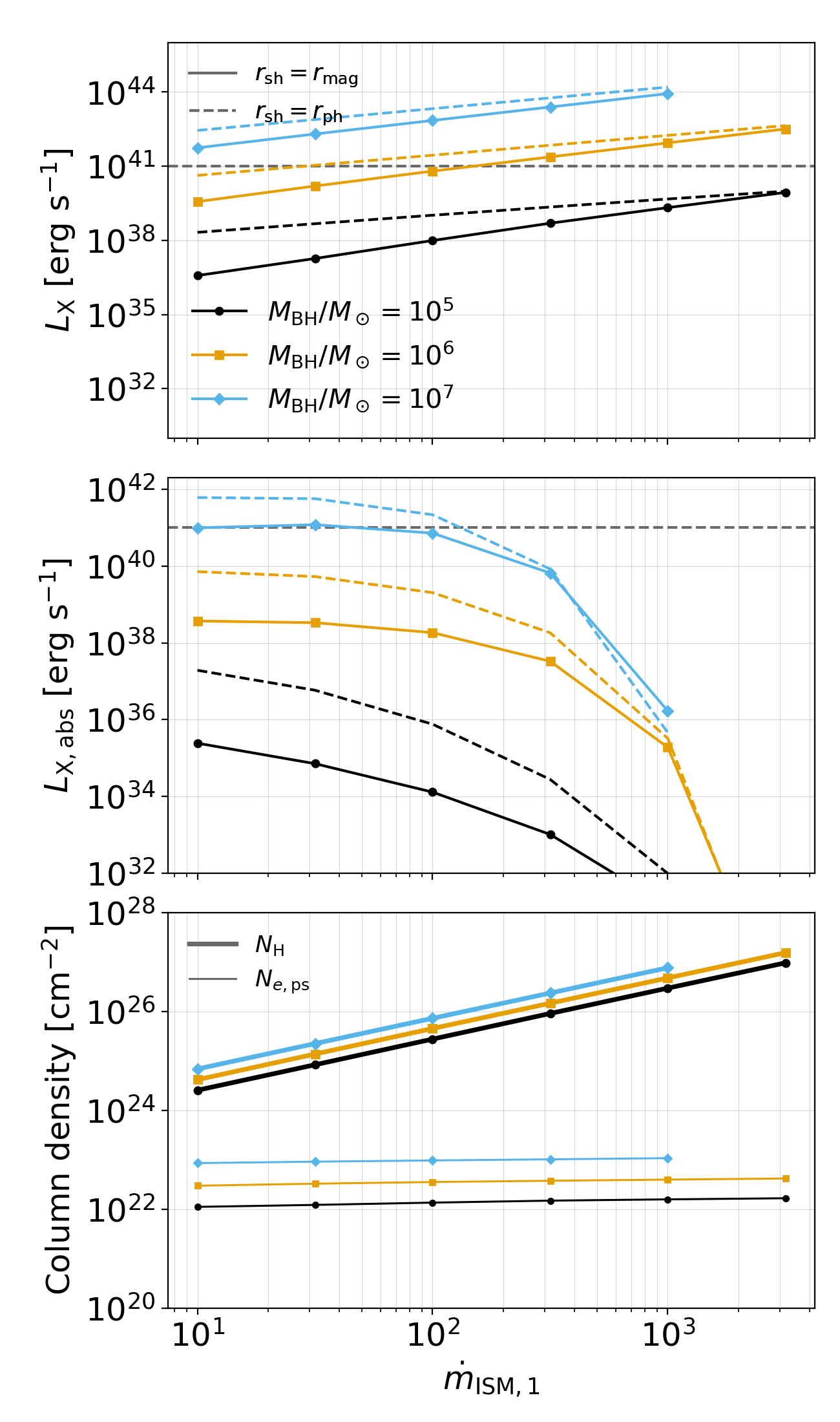}
\caption{The dependence of accretion-origin X-ray luminosity $L_{\rm X}$ on the normalized accretion rate $\dot{m}_{\rm ISM,1}$ for various black hole masses $M_{\rm BH}$. The top panel shows the intrinsic X-ray luminosity produced in the post-shock region, while the middle panel displays the X-ray luminosity attenuated by photoelectric absorption. In both panels, solid and dashed curves represent results with and without magnetospheres. Horizontal dashed lines denote the value of $L_{\rm X}=10^{41}~{\rm erg~s^{-1}}$. The X-ray luminosity is calculated from the model spectral luminosity in the photon energy range of $2 \text{-} 10$~keV. The bottom panel presents the column densities: $N_{\rm H}$ (solid) and $N_{\rm e,ps}$ (thin solid). $N_{\rm H}$ is used for the attenuation calculation.\label{fig:shock_Lx_mdot}}
\end{figure}

The top panel of Figure~\ref{fig:shock_Lx_mdot} shows the relationship between the intrinsic X-ray luminosity and the normalized accretion rate $\dot{m}_{\rm ISM,1}$. The intrinsic X-ray luminosity monotonically increases with $\dot{m}_{\rm ISM,1}$. For the cases corresponding to $M_{\rm BH}/M_\odot=10^6$ and $10^7$, the intrinsic X-ray luminosity is found to exceed the observational upper limit, $10^{41}~{\rm erg~s^{-1}}$ \citep{Lin_2026}. In contrast, the X-ray luminosity for $M_{\rm BH}/M_\odot=10^5$ remains below this limit due to low radiative efficiency of X-rays (Figure~\ref{fig:shock_Tps_Lx_over_Lacc}).

Since intrinsic X-rays are expected to be attenuated by accretion flows, we incorporate photoelectric absorption \citep{Osterbrock2006agna.book} to better assess the possibility of observational detection. We approximate the column density using $N_{\rm H}=\rho_{\rm acc,mag}r_{\rm ref}/m_{\rm p}$, where $m_{\rm p}$ represents the proton mass and $r_{\rm ref}$ denotes a typical spatial scale for density variation within the accretion flow. For this analysis, we assume that the accreting flow is mostly neutral. We adopt $r_{\rm ref}=r_{\rm mag}$ and $r_{\rm ph}$ for the cases with and without magnetospheres, respectively.

The middle panel of Figure~\ref{fig:shock_Lx_mdot} illustrates how the attenuated X-ray luminosity depends on $\dot{m}_{\rm ISM,1}$. When attenuation is accounted for, we find that the X-ray luminosities for $M_{\rm BH}/M_\odot=10^5$--$10^{7}$ are $\lesssim 10^{41}~{\rm erg~s^{-1}}$. Consequently, the X-ray luminosities no longer exhibit a monotonic increase because photoelectric absorption becomes more significant at higher values of $\dot{m}_{\rm ISM,1}$.
Furthermore, models with lower shock temperatures are subject to more severe photoelectric absorption (for example, comparing results with and without magnetospheres). This effect is a direct consequence of the energy dependency of the cross section.

In our spherical accretion model, the BHE is surrounded by accreting flows that contain a hot thin shell of post-shock gas (see also Figure~\ref{fig:lrd-illust}). The bottom panel of Figure~\ref{fig:shock_Lx_mdot} displays the column densities for these two layers: the neutral accretion flow and the fully ionized post-shock region. We find that the electron column density in the fully ionized post-shock region, $N_{\rm e,ps}$, is smaller than $10^{24}~{\rm cm^{-2}}$, suggesting that this layer is optically thin to Thomson scattering. Although ionizing photons from the post-shock gas will ionize a fraction of the accreting flows, the resulting electron column density outside the accretion shock remains smaller than that within the post-shock region. Consequently, radiation emerging from the BHE will be largely unaffected by the post-shock gas itself. However, the emerging radiation may still be altered by neutral accreting flows through resonance and Raman scattering \citep{Chang_2026_Balmer}.

\subsubsection{Magnetic-origin X-rays: Magnetically Heated Corona} \label{subsec:corona}
Convective stars, such as the Sun and CTTSs, exhibit X-ray emitting coronae \citep[e.g.][]{Gudel2004A&ARv}. If BHEs are also convective and magnetized via dynamo activities, they may form hot, X-ray emitting coronae (Figure~\ref{fig:lrd-illust}). To validate our BHE-as-a-protostar model, we must investigate the expected coronal X-ray luminosity.

We adopt a general concept of solar/stellar coronal heating based on energy balance in a steady atmosphere \citep{Rosner1978ApJ, Takasao2020ApJ}. Suppose that a hot, X-ray emitting corona with temperature $T_{\rm c} (> 10^6~{\rm K})$ forms in response to magnetic heating. Since this corona resides above the cooler underlying atmosphere, thermal conduction by free electrons transports energy toward the low atmosphere. This lower region receives the thermal energy and subsequently releases it as radiation. By denoting the thermal conduction flux and the radiative flux as $F_{\rm c}$ and $F_{\rm r}$, respectively, we obtain the following steady-state relation for the corona:
\begin{align}
    F_c \approx F_r.
\end{align}
The thermal conduction flux $F_c$ is approximated as
\begin{align}
    F_{\rm c}\sim \kappa_0 \frac{T_{\rm c}^{7/2}}{l_T},
\end{align}
where we assume Spitzer-type conductivity and $\kappa_0\approx 10^{-6}$ \citep{Spitzer1956}. Here, $l_T$ denotes the spatial scale of the temperature gradient (approximately the distance between the location of the temperature maximum and the base of the corona along a field line). The radiative flux is estimated as
\begin{align}
    F_{\rm r}\approx n_{\rm b}^2 \Lambda(T_{\rm b}) l_n,
\end{align}
where $n_{\rm b}$ and $T_{\rm b}$ represent the number density and temperature at the coronal base, respectively. $l_n$ denotes the spatial scale of the density gradient in the radial direction.

The X-ray luminosity $L_{\rm X}$ of the coronal base is estimated as follows:
\begin{align}
    L_{\rm X}\approx \int n_{\rm b}^2\Lambda(T_{\rm b})dV,
\end{align}
where the integrated volume corresponds to the volume of the low corona. When $l_n \ll r_{\rm ph}$, we can estimate this volume using a thin shell approximation, yielding $4\pi r_{\rm ph}^2 l_{n}$. Combining this relation with the energy balance yields the following relationship:
\begin{align}
    L_{\rm X}&\sim n_{\rm b}^2 \Lambda(T_{\rm b})\cdot 4\pi r_{\rm ph}^2 l_n \nonumber \\
    & \approx 4\pi \kappa_0 r_{\rm ph}^2 l_{T}^{-1} T_{\rm c}^{7/2}. \label{eq:corona_Lx}
\end{align}
As a measure of $l_T$, we take the pressure scale height $H$ for the coronal temperature:
\begin{align}
    H &= \frac{k_{\rm B}T_{\rm c}}{\mu m_{\rm p} g_{\rm ph}} \nonumber \\
    &\approx 1.2\times 10^{16}~{\rm cm}~m_{\rm BH,7}^{-1}r_{\rm ph,17}^{2}T_{\rm c,7},\label{eq:corona_H} 
\end{align}
where $k_{\rm B}$ and $m_{\rm p}$ are the Boltzmann constant and the proton mass, respectively. This result confirms that $H \ll r_{\rm ph}$ for $m_{\rm BH,7}\sim 1$, thereby validating our approximation of the thin shell volume for the typical black hole mass. Using a nondimensional constant $f_{H}$, we express $l_T$ as
\begin{align}
    l_T = f_H H. \label{eq:corona_l_T}
\end{align}
For the solar corona, $f_H\sim \mathcal{O}(1)$--$\mathcal{O}(10)$. Using Equations~(\ref{eq:corona_Lx}), (\ref{eq:corona_H}) and (\ref{eq:corona_l_T}), we derive the relation for the coronal X-ray luminosity as a function of $M_{\rm BH}, T_{\rm c}$ and $f_H$:
\begin{align}
    L_{\rm X} &\sim \frac{4\pi \kappa_0 \mu m_{\rm p}G}{k_{\rm B}}M_{\rm BH}f_H^{-1}T_{\rm c}^{5/2} \nonumber \\
    &\approx 3.2\times 10^{37}~{\rm erg~s^{-1}}~m_{\rm BH,7}f_{H,0}^{-1}T_{\rm c,7}^{5/2}\label{eq:corona_Lx_numerics}
\end{align}

\begin{figure}
\includegraphics[width=1.0\columnwidth]{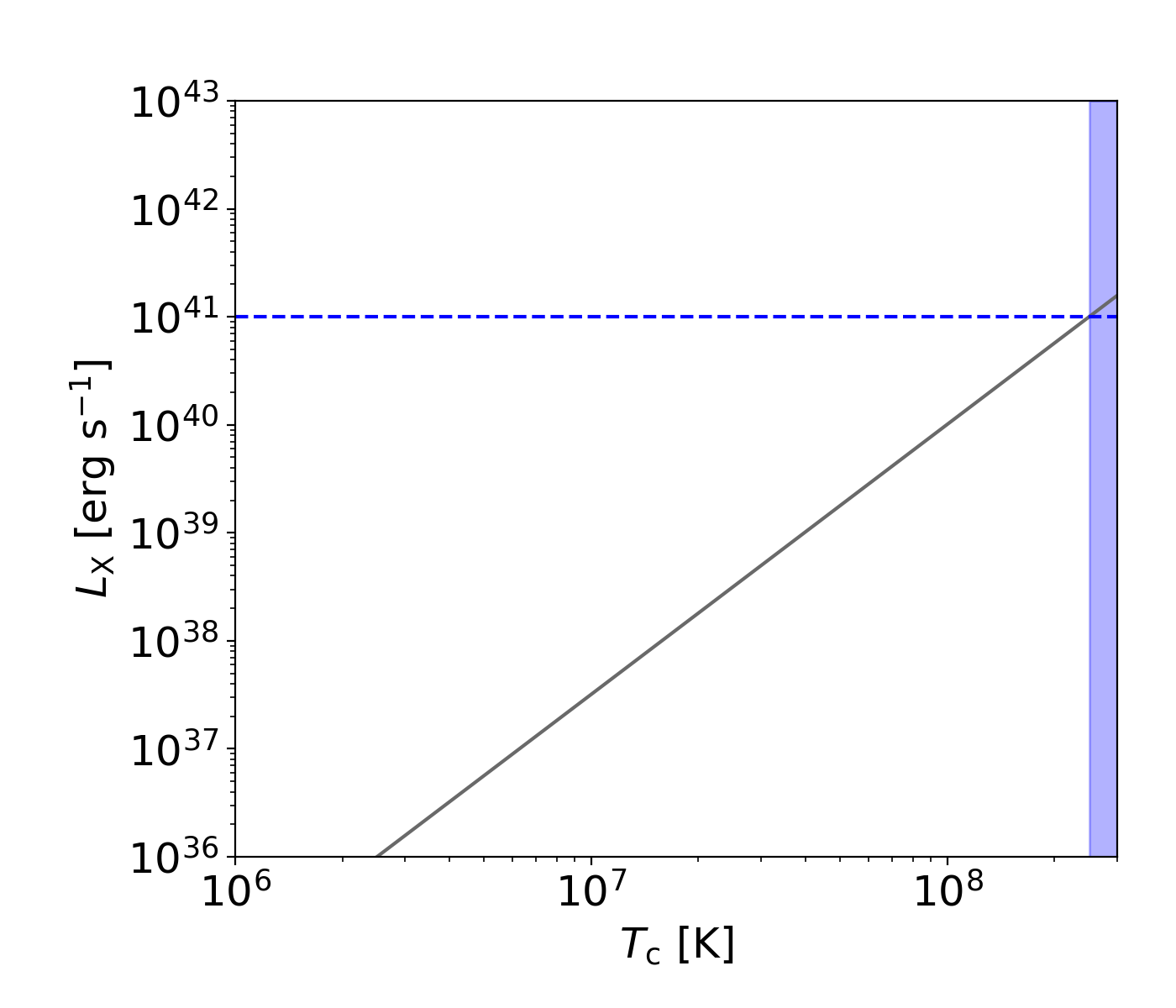}
\caption{Coronal X-ray luminosity as a function of the coronal temperature $T_{\rm c}$ for $M_{\rm BH}=10^7M_\odot$. The horizontal dashed line marks an observational upper limit, $10^{41}~{\rm erg~s^{-1}}$, and the shaded region is excluded by this constraint. \label{fig:corona_Lx_vs_Tc}}
\end{figure}

Figure~\ref{fig:corona_Lx_vs_Tc} shows $L_{\rm X}$ as a function of the coronal temperature $T_{\rm c}$ for the case of $f_{\rm H}$ and $M_{\rm BH}=10^{7}M_\odot$. Considering the observational constraint of the X-ray luminosity ($10^{41}~{\rm erg~s^{-1}}$), the coronal temperature must be constrained to $T_{\rm c}\le T_{\rm c,up}$, where $T_{\rm c,up}$ is estimated as:
\begin{align}
    T_{\rm c,up}\sim 3\times 10^8~{\rm K}~f_{H,0}^{2/5}L_{\rm X,41}^{2/5}m_{\rm BH,7}^{-2/5} \label{eq:upper-lim-Tc}
\end{align}
We note that a higher $T_{\rm c}$ is allowed for a smaller $M_{\rm BH}$, according to the relation~(\ref{eq:corona_Lx_numerics}).

The mechanism determining the coronal temperature $T_{\rm c}$ remains uncertain due to ambiguities regarding the detailed energy injection processes and the stratified atmospheric structure. Nevertheless, magnetically active low-mass stars, including CTTSs, commonly show $T_{\rm c} \lesssim \text{a few}~10^{7}~{\rm K}$ \citep[except for the phase of flares driven by magnetic reconnection. See e.g.,][]{Preibisch2005ApJS}. The apparent upper limit may be constrained by the strong dependence of thermal conductivity on $T_{\rm c}$. Furthermore, since this apparent upper limit is comparable to the virial temperature, gravitational bounds represent another key factor. For BHEs, the virial temperature is estimated as
\begin{align}
    T_{\rm vir}&=\frac{\mu m_{\rm p}}{2k_{\rm B}}\frac{GM_{\rm BH}}{r_{\rm ph}} \nonumber \\
&\approx 4.0\times 10^7~{\rm K}~m_{\rm BH,7}r_{\rm ph,17}^{-1},
\end{align}
which is lower than the upper limit of the allowed coronal temperature for $M_{\rm BH}=10^{7}M_\odot$ ($\sim 3\times  10^{8}~{\rm K}$, see Figure~\ref{fig:corona_Lx_vs_Tc}). For BHEs with smaller $M_{\rm BH}$, the virial temperature decreases while the upper limit increases (Equation~(\ref{eq:upper-lim-Tc})). As a result, we expect that BHE coronae will not produce strong, detectable X-rays even if they are present.

\section{Discussion} \label{sec:discussion}

\subsection{Accretion-origin Emissions}
After passing through an accretion shock, accretion flows cool from a temperature of $10^7\text{-}10^8$~K. This cooling process suggests the formation of hot emission lines of hydrogen and helium, which are generally not detected in LRDs. Our post-shock model indicates that while these hot emission lines will form, they will be significantly weaker than the BHE photospheric emission (see Figures~\ref{fig:shock_spec_lum} and \ref{fig:sed_obs_model}). As a result, our BHE model is consistent with the non-detection of such hot emission lines.

The modeled spectral luminosity's UV-to-optical ratio is significantly smaller than typical observed values (Section~\ref{subsec:accretion_shock}). For the case of $(m_{\rm BH,7},\dot{m}_{\rm ISM,1})=(0.1,10)$ (Figure~\ref{fig:shock_spec_lum}), the total accretion luminosity at the magnetospheric radius is only $\sim 5\times 10^{41}~{\rm erg~s^{-1}}$. In contrast, the BHE photospheric luminosity ($\sim 10^{44}~{\rm erg~s^{-1}}$) is much larger. Consequently, a small ratio for shock-origin UV to optical emission is expected. Our model therefore requires additional sources of UV radiation to account for the full SEDs of LRDs, such as starburst activity \citep{Inayoshi_2025d}.

In Section~\ref{subsec:accretion_shock}, we focused only on the optically thin post-shock layer. The structure of the cooler and denser layer below this region is reserved for future investigation. However, in cases of stellar accretion within CTTSs, an accretion-heated, optically thick layer forms beneath the optically thin post-shock layer \citep{Hartmann2016ARA&A}. The effective temperature of this heated layer, $T_{\rm eff,shock}$, can be estimated from energy balance: $\sigma_{\rm SB}T_{\rm eff,shock}^4\sim \rho_{\rm acc}v_{\rm acc}^3/2$, where $\sigma_{\rm SB}$ is the Stefan-Boltzmann constant \citep[e.g.][]{Romanova2004ApJ}, and $\rho_{\rm acc}$ and $v_{\rm acc}$ denote the density and speed of the accretion flow. Applying this picture to the accretion shock around the BHE at $r_{\rm mag}$ yields the following effective temperature for the shocked gas:
\begin{align}
    T_{\rm eff,shock} &\sim \left(\frac{\rho_{\rm acc,mag}v_{\rm ff,mag}^3}{2\sigma_{\rm SB}}\right)^{1/4} \nonumber \\
    &\approx 2.5\times 10^2~{\rm K}~\tilde{\kappa}^{3/14}m_{\rm BH,7}^{1/7}\dot{m}_{\rm ISM,1}^{13/28}r_{\rm ph,17}^{-6/7} \nonumber\\
    &\approx 1.3\times 10^3~{\rm K}~m_{\rm BH,7}^{1/4}\dot{m}_{\rm ISM,2}^{1/4}r_{\rm mag,17}^{-3/4},
\end{align}
where $\rho_{\rm acc,mag}=\dot{M}_{\rm ISM}/(4\pi r_{\rm mag}^2v_{\rm acc,mag})$. 
Since this calculated temperature is significantly smaller than the observationally inferred photospheric temperature ($\sim 5000$~K; see, e.g., \citet{Lin_2026}), we consider that the shocked gas cannot entirely cover the BHE by forming an optically thick shell. This picture aligns with our argument regarding the possible origin of BLRs (Sections~\ref{subsec:general} and \ref{subsec:rotation}). Namely, if the shocked gas does not accumulate at the magnetospheric radius, either because it penetrates into the magnetosphere (likely in the non-propeller regime) or because it is disturbed by outflows (likely in the propeller regime), an optically thick shell will not form.

Since the typical temperature of $T_{\rm eff,shock}\sim 1.3\times 10^3$~K is comparable to the dust sublimation temperature, we expect potential confusion with genuine dust emission. Observations have shown that a fraction of LRDs, as well as stacked SEDs, exhibit an infrared excess \citep{Lyu_2024,Delvecchio_2025,Lin_2026}, which has been conventionally interpreted as evidence for warm dust. However, our BHE model suggests that this infrared excess may instead be produced by optically thick gas in the post-shock region. If this is the case, the magnitude of the excess could provide a constraint on the filling factor of the optically thick gas.

\subsection{Role of Observations in Understanding Magnetic Properties of Giant Stars}

Giant stars and BHEs share several fundamental physical characteristics, including convective interiors capable of generating magnetic fields through dynamo action. Indeed, spectroscopic observations have directly detected stellar magnetic fields in a fraction of giant stars \citep[e.g.,][]{Vlemmings2005A&A,Auriere2015A&A,Tessore2017A&A}. In addition, \citet{Lin_2026} identified absorption features of Na~D, K, and Ca~II in LRDs, resembling spectral features commonly observed in the atmospheres of cool stars. Although these absorption features do not directly indicate the presence of magnetic fields, future spectroscopic studies may help clarify whether they are related to magnetic activity and field properties in BHEs.

Investigating the magnetic properties of LRDs is crucial for assessing the validity of our magnetized BHE model. Furthermore, if LRDs represent precursors to AGNs, characterizing their magnetic fields may provide insight into the origin of magnetism during the AGN phase. Although magnetic fields in AGNs are believed to play a central role in launching jets, their ultimate origin remains uncertain. Existing scenarios include dynamo amplification within accretion disks and the advection of ambient magnetic fields \citep[e.g.,][]{Hawley2015SSRv}. We propose that magnetic fields generated by dynamo action within BHEs represent an additional channel for the origin of AGN magnetic fields.

Considering the existing stellar and LRD observations, investigating the relationship between stellar magnetic properties and atmospheric absorption lines (such as Na D and Ca~II) offers a promising diagnostic tool for testing our magnetized BHE model. Such links have been extensively studied in solar-type stars; we will briefly introduce this approach to motivate similar observational studies in giant stars and LRDs.

In solar-type stars, Ca II absorption features originate in the chromosphere. Stronger stellar magnetic fields correlate with increased absorption core intensity due to enhanced chromospheric heating. Solar observations have established empirical relations between the depth of these absorption lines and the averaged stellar magnetic field strength, enabling us to infer the average field strength of unresolved solar-type stars \citep[e.g.][]{Schrijver1989ApJ, Notsu2015PASJ}. Furthermore, this average field strength allows for the computation of the total magnetic flux. This total magnetic flux is a valuable parameter used to estimate the coronal X-ray luminosity \citep{Pevtsov2003ApJ}.

Dwarf star observations have developed useful techniques for studying stellar magnetic properties. Applying these methods to LRDs will open a new avenue for testing the hypothesis of magnetized BHEs. Additionally, it is important to observationally examine whether empirical relations derived from cool dwarfs apply equally to giant stars and BHEs. Significant physical differences among cool dwarfs, giants, and BHEs, particularly in surface gravity and the role of radiation pressure, must be taken into account in this comparison. For detailed information regarding magnetic field observations and spectroscopic features in cool stars, we refer readers to \citet{Reiners2012LRSP} and \citet{Linsky2017ARA&A}.

\subsection{Some Remarks on BHE Spins}
Section~\ref{subsec:rotation} highlights that the BHE spin is crucial for interpreting the Doppler broadening of emission lines. BHEs can acquire significant angular momentum from accreting flows. If these objects contract in response to Kelvin-Helmholtz contraction without losing their angular momentum, they will spin up and potentially enter the propeller regime. Therefore, a detailed understanding of BHE spin evolution is critical for correctly interpreting observational data. Furthermore, if BHEs are subject to non-spherical disk accretion, knowledge of star-disk interaction in solar-mass protostars will become particularly useful for modeling the spin-up/down processes of BHEs (see \citet{Takasao2025ApJ_a} and references therein).

We estimate the critical conditions separating the propeller and non-propeller regimes. This condition is expressed as $r_{\rm cor}=r_{\rm mag}$. Using the relation given by $r_{\rm cor}=f_{\rm rot}^{-2/3}r_{\rm ph}$ (Equations~(\ref{eq:r_cor}) and (\ref{eq:definition_f_rot})), we find the critical value of $f_{\rm rot}$ to be 
\begin{align}
    f_{\rm rot,crit}=\left(\frac{r_{\rm mag}}{r_{\rm ph}}\right)^{-3/2}\approx 0.11\tilde{\kappa}^{3/7}m_{\rm BH,7}^{-3/14}\dot{m}_{\rm ISM,1}^{3/7}r_{\rm ph,17}^{-3/14},
\end{align}
where we have used Equation~(\ref{eq:rmag_rph}).
If $f_{\rm rot}>f_{\rm rot,crit}$, the BHE will be in the propeller regime.

BHEs in the propeller regime are expected to produce outflows, whose signatures may be imprinted on spectral line profiles as P-Cygni features. However, MHD simulations \citep{Toropina2005MmSAI} suggest that these outflows might be confined to a limited region by the surrounding accretion flows. Consequently, the formation of failed outflows around the equatorial plane could cause the accretion shock to become oblate and potentially reduce the overall accretion rate. Investigating these complex effects requires future dedicated studies.

\section{Summary} \label{sec:summary}
We present a theoretical model for LRDs, assuming that a black hole is embedded within a dense BHE behaving like a convective protostar near the Hayashi-limit. This framework incorporates free-falling spherical accretion at high rates exceeding the Eddington value. By considering magnetic field effects, we investigate the dynamical and spectral properties of this BHE-as-a-protostar model. The main findings are as follows:

\begin{itemize}
    \item If BHEs possess surface magnetic fields similar to the thermal equipartition field $B_{\rm eq}$, they develop a magnetosphere across a wide range of black hole masses and accretion rates. In particular, for BHEs with $M_{\rm BH}\le 10^{7}~M_{\odot}$, this magnetosphere remains uncollapsed even at very high accretion rates (see Section~\ref{subsec:formulation} and Figure~\ref{fig:rmag_rph}).
    \item Our BHE model proposes a new interpretation of broad emission lines in LRDs (Section~\ref{subsec:rotation}). The model predicts that Doppler broadening arises from plasma clumps corotating with the BHE magnetosphere, contrasting with the standard AGN picture where line broadening is typically attributed to the virial motion of gas clumps. The rotating magnetized BHE model can produce Doppler components corresponding to a few thousand ${\rm km~s^{-1}}$. The line profile shaped by Doppler effects intrinsically exhibits a double-peaked structure. When additional broadening due to electron scattering is included, this double-peaked structure becomes smoothed, and the resulting profile can be fitted by a combination of a Gaussian-like core and an exponential tail.
    \item If LRDs are magnetized, rotating BHEs, conventional black hole mass estimation based on the virial relation may yield erroneous results (Figure~\ref{fig:FWHM-L5100}). We suggest that for LRDs exhibiting $\textrm{FWHM}_{\rm Doppler}\gtrsim 10^3~{\rm km~s^{-1}}$, the actual mass could be smaller by an order of magnitude or more. Conversely, for LRDs with $\textrm{FWHM}_{\rm Doppler}\lesssim 3\times 10^2~{\rm km~s^{-1}}$, the actual mass could be larger by a factor of three or more.
    \item Our BHE model accounts for the observed absence or weakness of X-rays in LRDs (Section~\ref{subsec:X-ray}). Within this framework, two potential sources contribute to X-ray emission: (1) the post-shock region of the accretion shock (Section~\ref{subsec:X-ray-shock}) and (2) the magnetically heated corona (Section~\ref{subsec:corona}). We evaluate the X-ray luminosities from these sources and find that they are lower than $10^{41}~{\rm erg~s^{-1}}$ over a wide parameter range, which is consistent with observationally inferred upper limits. Furthermore, regarding the accretion-origin emission, we show that BHE magnetic fields reduce the X-ray luminosity by lowering the shock temperature (e.g., Figures~\ref{fig:shock_Tps_Lx_over_Lacc} and \ref{fig:shock_Lx_mdot}).
\end{itemize}

\begin{acknowledgments}
We greatly thank Jorryt Matthee and Rohan Naidu for constructive discussions at the conference, Charting Cosmic Dawn In Copenhagen 2026. 
S.T. was supported by JSPS KAKENHI grant Nos. JP22KK0043, JP21H04487 and JP26K00769.
K.I. acknowledges support from the National Natural Science Foundation of China (12573015, W2532003), the Beijing Natural Science Foundation (IS25003), and the China Manned Space Program (CMS-CSST-2025-A09).

\end{acknowledgments}

%

\software{astropy}


\appendix

\section{Surface Magnetic Field Strength of Giant Stars} \label{app:bfield_giant}
To examine the ratio of observed field strength to the thermal equipartition field $B_{\rm eq}$ in giant stars, we utilize data from magnetically active giants reported by \citet{Auriere2015A&A}. This dataset includes Zeeman measurements of the maximum longitudinal magnetic field strength during observations, $|B_{l}|_{\rm max}$. We compute $B_{\rm eq}$ using observationally estimated surface gravity, effective temperature, and the opacity model (see Section~\ref{subsec:general}). Furthermore, we incorporate the S-index provided by \citet{Auriere2015A&A}, which measures magnetic activity based on chromospheric Ca II emissions; a larger S-index indicates a higher level of magnetic activity.

\begin{figure}
\plotone{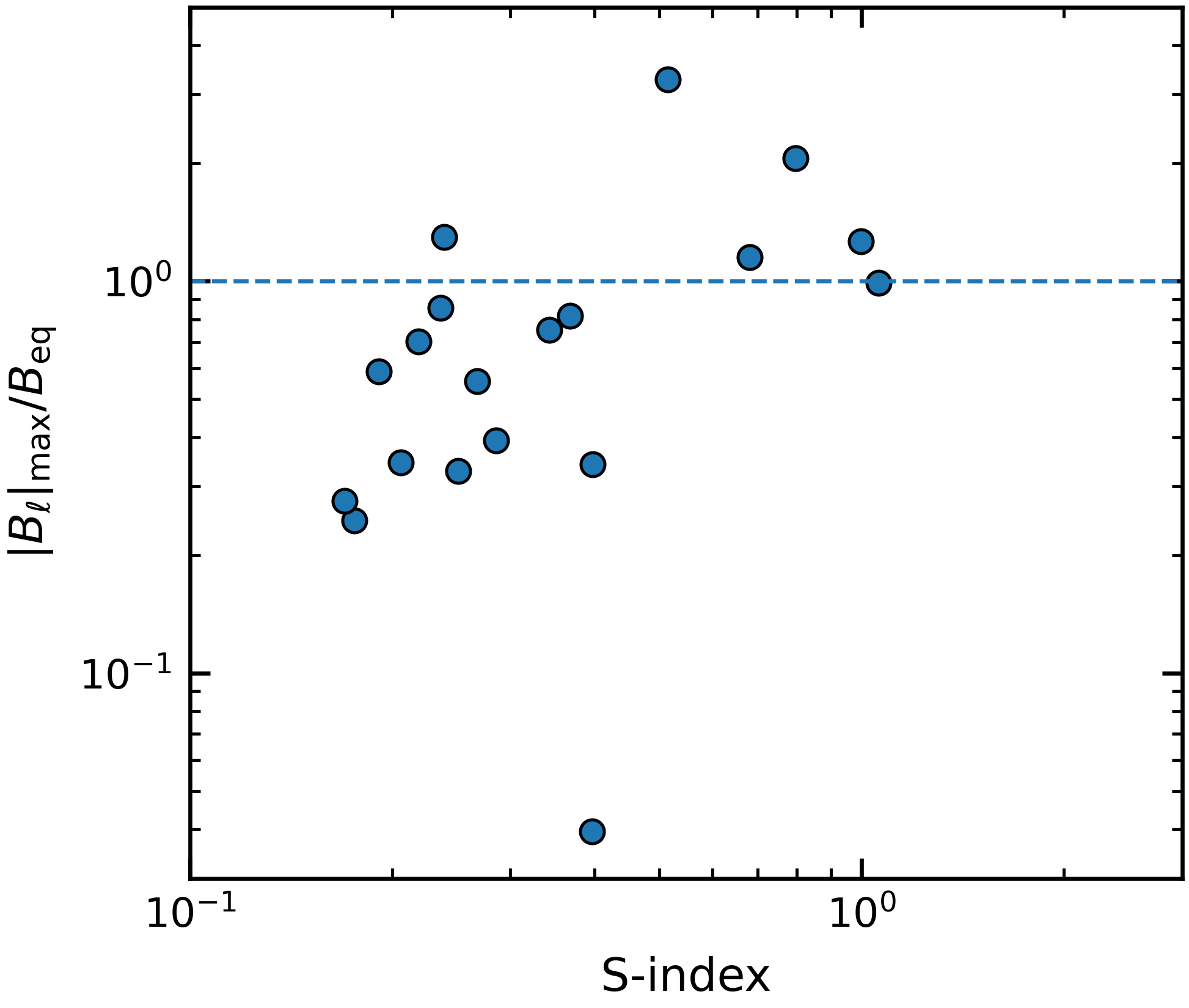}
\caption{Surface magnetic field strength of giant stars normalized by their thermal equipartition field, $B_{\rm eq}$. The data are sourced from \citet{Auriere2015A&A}. The horizontal dashed line denotes the ratio of unity. \label{fig:B_giant_obs}}
\end{figure}

Figure~\ref{fig:B_giant_obs} shows the normalized field strength, $|B_{l}|_{\rm max}/B_{\rm eq}$, plotted against the S-index. The data reveal that most samples exhibit $|B_{l}|_{\rm max}/B_{\rm eq}\sim 1$, which motivates us to assume $B_{\rm surf}\sim B_{\rm eq}$ in Section~\ref{sec:mag}.

\section{Fitting Formula of opacity}
Figure~\ref{fig:opacity} suggests that the tabulated Rosseland mean opacity can be fitted by a power law function below 6000~K. Given that the typical range of the photospheric temperature is $\lesssim 6000$~K, we derive a fitting function for comparison with the fit presented in \citet{Begelman2008MNRAS}. The resulting fitting function, only valid within the range of $10^{-16}~{\rm g~cm^{-3}}\le \rho \le 10^{-12}~{\rm g~cm^{-3}}$ and $3000~{\rm K}\le T \le 6000~{\rm K}$, is obtained as follows:
\begin{align}
    \kappa(\rho,T) &\approx 4.26\times 10^{-2}~{\rm cm^2~g^{-1}}\nonumber \\
    &\times \left(\frac{\rho}{10^{-15}~{\rm g~cm^{-3}}}\right)^{q_{\rm d}} \left(\frac{T}{5000~{\rm K}}\right)^{q_{\rm T}},\label{eq:opacity_fit}
\end{align}
where $q_{\rm d}=-0.37$ and $q_{\rm T}=16.76$. The typical scatter and worst error for this fit are approximately 30\% and 80\%, respectively. This function exhibits a stronger dependence on temperature than the fit provided by \citet{Begelman2008MNRAS}. To avoid error propagation, we did not use this fit to derive $r_{\rm ph}$ and $T_{\rm ph}$. Although one might calculate scaling relations of other quantities by combining this fit with the scalings derived from $r_{\rm ph}$ and $T_{\rm ph}$, we caution that the opacity's very strong temperature dependence could cause significant deviations from actual values. Therefore, we avoid deriving such scaling relations by combining these variables.


\bibliography{sample701}{}

\begin{thebibliography}{}
\expandafter\ifx\csname natexlab\endcsname\relax\def\natexlab#1{#1}\fi
\providecommand{\url}[1]{\href{#1}{#1}}
\providecommand{\dodoi}[1]{doi:~\href{http://doi.org/#1}{\nolinkurl{#1}}}
\providecommand{\doeprint}[1]{\href{http://ascl.net/#1}{\nolinkurl{http://ascl.net/#1}}}
\providecommand{\doarXiv}[1]{\href{https://arxiv.org/abs/#1}{\nolinkurl{https://arxiv.org/abs/#1}}}

\bibitem[{Y. {Aoyama} {et~al.}(2018){Aoyama}, {Ikoma}, \& {Tanigawa}}]{Aoyama2018ApJ}
{Aoyama}, Y., {Ikoma}, M., \& {Tanigawa}, T. 2018, \bibinfo{title}{{Theoretical Model of Hydrogen Line Emission from Accreting Gas Giants},} \apj, 866, 84, \dodoi{10.3847/1538-4357/aadc11}

\bibitem[{Y. {Asada} {et~al.}(2026){Asada}, {Inayoshi}, {Fei}, {Fujimoto}, \& {Willott}}]{Asada_2026}
{Asada}, Y., {Inayoshi}, K., {Fei}, Q., {Fujimoto}, S., \& {Willott}, C. 2026, \bibinfo{title}{{Origins of the UV continuum and Balmer emission lines in Little Red Dots: observational validation of dense gas envelope models enshrouding the AGN},} arXiv e-prints, arXiv:2601.10573, \dodoi{10.48550/arXiv.2601.10573}

\bibitem[{M. {Auri{\`e}re} {et~al.}(2015){Auri{\`e}re}, {Konstantinova-Antova}, {Charbonnel}, {Wade}, {Tsvetkova}, {Petit}, {Dintrans}, {Drake}, {Decressin}, {Lagarde}, {Donati}, {Roudier}, {Ligni{\`e}res}, {Schr{\"o}der}, {Landstreet}, {L{\`e}bre}, {Weiss}, \& {Zahn}}]{Auriere2015A&A}
{Auri{\`e}re}, M., {Konstantinova-Antova}, R., {Charbonnel}, C., {et~al.} 2015, \bibinfo{title}{{The magnetic fields at the surface of active single G-K giants},} \aap, 574, A90, \dodoi{10.1051/0004-6361/201424579}

\bibitem[{M.~C. {Begelman} \& J. {Dexter}(2026){Begelman} \& {Dexter}}]{Begelman2026ApJ}
{Begelman}, M.~C., \& {Dexter}, J. 2026, \bibinfo{title}{{Little Red Dots as Late-stage Quasi-stars},} \apj, 996, 48, \dodoi{10.3847/1538-4357/ae274a}

\bibitem[{M.~C. {Begelman} {et~al.}(2008){Begelman}, {Rossi}, \& {Armitage}}]{Begelman2008MNRAS}
{Begelman}, M.~C., {Rossi}, E.~M., \& {Armitage}, P.~J. 2008, \bibinfo{title}{{Quasi-stars: accreting black holes inside massive envelopes},} \mnras, 387, 1649, \dodoi{10.1111/j.1365-2966.2008.13344.x}

\bibitem[{M.~C. {Bentz} {et~al.}(2013){Bentz}, {Denney}, {Grier}, {Barth}, {Peterson}, {Vestergaard}, {Bennert}, {Canalizo}, {De Rosa}, {Filippenko}, {Gates}, {Greene}, {Li}, {Malkan}, {Pogge}, {Stern}, {Treu}, \& {Woo}}]{Benzt2013ApJ}
{Bentz}, M.~C., {Denney}, K.~D., {Grier}, C.~J., {et~al.} 2013, \bibinfo{title}{{The Low-luminosity End of the Radius-Luminosity Relationship for Active Galactic Nuclei},} \apj, 767, 149, \dodoi{10.1088/0004-637X/767/2/149}

\bibitem[{T.-Q. {Cang} {et~al.}(2020){Cang}, {Petit}, {Donati}, {Folsom}, {Jardine}, {Villarreal D'Angelo}, {Vidotto}, {Marsden}, {Gallet}, \& {Zaire}}]{Cang2020A&A}
{Cang}, T.-Q., {Petit}, P., {Donati}, J.-F., {et~al.} 2020, \bibinfo{title}{{Magnetic field and prominences of the young, solar-like, ultra-rapid rotator V530 Persei},} \aap, 643, A39, \dodoi{10.1051/0004-6361/202037693}

\bibitem[{S.-J. {Chang} {et~al.}(2026){Chang}, {Gronke}, {Matthee}, \& {Mason}}]{Chang_2026_Balmer}
{Chang}, S.-J., {Gronke}, M., {Matthee}, J., \& {Mason}, C. 2026, \bibinfo{title}{{Impact of resonance, Raman, and Thomson scattering on hydrogen line formation in Little Red Dots},} \mnras, 545, staf2131, \dodoi{10.1093/mnras/staf2131}

\bibitem[{A. {Comastri} {et~al.}(2025){Comastri}, {Lanzuisi}, {Vito}, {Marchesi}, {Brusa}, {Gilli}, {Juodzbalis}, {Maiolino}, {Mazzolari}, {Risaliti}, {Scholtz}, \& {Vignali}}]{Comastri_2025}
{Comastri}, A., {Lanzuisi}, G., {Vito}, F., {et~al.} 2025, \bibinfo{title}{{JWST-discovered AGN: evidence for heavy obscuration in the type-2 sample from the first stacked X-ray detection},} arXiv e-prints, arXiv:2510.00112, \dodoi{10.48550/arXiv.2510.00112}

\bibitem[{S. {Daley-Yates} \& M.~M. {Jardine}(2024){Daley-Yates} \& {Jardine}}]{Daley-Yates2024MNRAS}
{Daley-Yates}, S., \& {Jardine}, M.~M. 2024, \bibinfo{title}{{Simulating stellar coronal rain and slingshot prominences},} \mnras, 534, 621, \dodoi{10.1093/mnras/stae2131}

\bibitem[{A. {de Graaff} {et~al.}(2025{\natexlab{a}}){de Graaff}, {Rix}, {Naidu}, {Labb{\'e}}, {Wang}, {Leja}, {Matthee}, {Katz}, {Greene}, {Hviding}, {Baggen}, {Bezanson}, {Boogaard}, {Brammer}, {Dayal}, {van Dokkum}, {Goulding}, {Hirschmann}, {Maseda}, {McConachie}, {Miller}, {Nelson}, {Oesch}, {Setton}, {Shivaei}, {Weibel}, {Whitaker}, \& {Williams}}]{deGraaff_2025b}
{de Graaff}, A., {Rix}, H.-W., {Naidu}, R.~P., {et~al.} 2025{\natexlab{a}}, \bibinfo{title}{{A remarkable ruby: Absorption in dense gas, rather than evolved stars, drives the extreme Balmer break of a little red dot at z = 3.5},} \aap, 701, A168, \dodoi{10.1051/0004-6361/202554681}

\bibitem[{A. {de Graaff} {et~al.}(2025{\natexlab{b}}){de Graaff}, {Hviding}, {Naidu}, {Greene}, {Miller}, {Leja}, {Matthee}, {Brammer}, {Katz}, {Bezanson}, {Boogaard}, {Bose}, {Chisholm}, {Cleri}, {Dayal}, {Feldmann}, {Fudamoto}, {Fujimoto}, {Furtak}, {Glazebrook}, {Gottumukkala}, {Heintz}, {Kokorev}, {Labbe}, {Maseda}, {McConachie}, {Nanayakkara}, {Nelson}, {Nowaczyk}, {Oesch}, {Rix}, {Setton}, {Torralba}, {Walter}, {Wang}, {Weibel}, \& {van der Wel}}]{deGraaff_2025c}
{de Graaff}, A., {Hviding}, R.~E., {Naidu}, R.~P., {et~al.} 2025{\natexlab{b}}, \bibinfo{title}{{Little Red Dots host Black Hole Stars: A unified family of gas-reddened AGN revealed by JWST/NIRSpec spectroscopy},} arXiv e-prints, arXiv:2511.21820, \dodoi{10.48550/arXiv.2511.21820}

\bibitem[{G. {Del Zanna} {et~al.}(2021){Del Zanna}, {Dere}, {Young}, \& {Landi}}]{DelZanna2021ApJ}
{Del Zanna}, G., {Dere}, K.~P., {Young}, P.~R., \& {Landi}, E. 2021, \bibinfo{title}{{CHIANTI{\textemdash}An Atomic Database for Emission Lines. XVI. Version 10, Further Extensions},} \apj, 909, 38, \dodoi{10.3847/1538-4357/abd8ce}

\bibitem[{I. {Delvecchio} {et~al.}(2025){Delvecchio}, {Daddi}, {Magnelli}, {Elbaz}, {Giavalisco}, {Traina}, {Lanzuisi}, {Akins}, {Belli}, {Casey}, {Gentile}, {Gruppioni}, {Pozzi}, \& {Zamorani}}]{Delvecchio_2025}
{Delvecchio}, I., {Daddi}, E., {Magnelli}, B., {et~al.} 2025, \bibinfo{title}{{AGN-heated dust revealed in ``Little Red Dots''},} arXiv e-prints, arXiv:2509.07100, \dodoi{10.48550/arXiv.2509.07100}

\bibitem[{R.~F. {Elsner} \& F.~K. {Lamb}(1977){Elsner} \& {Lamb}}]{Elsner1977ApJ}
{Elsner}, R.~F., \& {Lamb}, F.~K. 1977, \bibinfo{title}{{Accretion by magnetic neutron stars. I. Magnetospheric structure and stability.},} \apj, 215, 897, \dodoi{10.1086/155427}

\bibitem[{L.~J. {Furtak} {et~al.}(2024){Furtak}, {Labb{\'e}}, {Zitrin}, {Greene}, {Dayal}, {Chemerynska}, {Kokorev}, {Miller}, {Goulding}, {de Graaff}, {Bezanson}, {Brammer}, {Cutler}, {Leja}, {Pan}, {Price}, {Wang}, {Weaver}, {Whitaker}, {Atek}, {Bogd{\'a}n}, {Charlot}, {Curtis-Lake}, {van Dokkum}, {Endsley}, {Feldmann}, {Fudamoto}, {Fujimoto}, {Glazebrook}, {Juneau}, {Marchesini}, {Maseda}, {Nelson}, {Oesch}, {Plat}, {Setton}, {Stark}, \& {Williams}}]{Furtak2024Natur}
{Furtak}, L.~J., {Labb{\'e}}, I., {Zitrin}, A., {et~al.} 2024, \bibinfo{title}{{A high black-hole-to-host mass ratio in a lensed AGN in the early Universe},} \nat, 628, 57, \dodoi{10.1038/s41586-024-07184-8}

\bibitem[{J.~E. {Greene} \& L.~C. {Ho}(2005){Greene} \& {Ho}}]{Greene2005ApJ}
{Greene}, J.~E., \& {Ho}, L.~C. 2005, \bibinfo{title}{{Estimating Black Hole Masses in Active Galaxies Using the H{\ensuremath{\alpha}} Emission Line},} \apj, 630, 122, \dodoi{10.1086/431897}

\bibitem[{J.~E. {Greene} {et~al.}(2024){Greene}, {Labbe}, {Goulding}, {Furtak}, {Chemerynska}, {Kokorev}, {Dayal}, {Volonteri}, {Williams}, {Wang}, {Setton}, {Burgasser}, {Bezanson}, {Atek}, {Brammer}, {Cutler}, {Feldmann}, {Fujimoto}, {Glazebrook}, {de Graaff}, {Khullar}, {Leja}, {Marchesini}, {Maseda}, {Matthee}, {Miller}, {Naidu}, {Nanayakkara}, {Oesch}, {Pan}, {Papovich}, {Price}, {van Dokkum}, {Weaver}, {Whitaker}, \& {Zitrin}}]{Greene_2024}
{Greene}, J.~E., {Labbe}, I., {Goulding}, A.~D., {et~al.} 2024, \bibinfo{title}{{UNCOVER Spectroscopy Confirms the Surprising Ubiquity of Active Galactic Nuclei in Red Sources at z > 5},} \apj, 964, 39, \dodoi{10.3847/1538-4357/ad1e5f}

\bibitem[{M. {G{\"u}del}(2004){G{\"u}del}}]{Gudel2004A&ARv}
{G{\"u}del}, M. 2004, \bibinfo{title}{{X-ray astronomy of stellar coronae},} \aapr, 12, 71, \dodoi{10.1007/s00159-004-0023-2}

\bibitem[{L. {Hartmann} {et~al.}(2016){Hartmann}, {Herczeg}, \& {Calvet}}]{Hartmann2016ARA&A}
{Hartmann}, L., {Herczeg}, G., \& {Calvet}, N. 2016, \bibinfo{title}{{Accretion onto Pre-Main-Sequence Stars},} \araa, 54, 135, \dodoi{10.1146/annurev-astro-081915-023347}

\bibitem[{J.~F. {Hawley} {et~al.}(2015){Hawley}, {Fendt}, {Hardcastle}, {Nokhrina}, \& {Tchekhovskoy}}]{Hawley2015SSRv}
{Hawley}, J.~F., {Fendt}, C., {Hardcastle}, M., {Nokhrina}, E., \& {Tchekhovskoy}, A. 2015, \bibinfo{title}{{Disks and Jets. Gravity, Rotation and Magnetic Fields},} \ssr, 191, 441, \dodoi{10.1007/s11214-015-0174-7}

\bibitem[{C. {Hayashi}(1961){Hayashi}}]{Hayashi1961PASJ_b}
{Hayashi}, C. 1961, \bibinfo{title}{{Stellar Evolution in Early Phases of Gravitational Contraction},} \pasj, 13, 450, \dodoi{10.1093/pasj/13.4.450}

\bibitem[{C. {Hayashi} \& R. {H{\={o}}shi}(1961){Hayashi} \& {H{\={o}}shi}}]{Hayashi1961PASJ_a}
{Hayashi}, C., \& {H{\={o}}shi}, R. 1961, \bibinfo{title}{{The Outer Envelope of Giant Stars with Surface Convection Zone},} \pasj, 13, 442, \dodoi{10.1093/pasj/13.4.442}

\bibitem[{R.~E. {Hviding} {et~al.}(2025){Hviding}, {de Graaff}, {Miller}, {Setton}, {Greene}, {Labb{\'e}}, {Brammer}, {Bezanson}, {Boogaard}, {Cleri}, {Leja}, {Maseda}, {McConachie}, {Matthee}, {Naidu}, {Oesch}, {Wang}, {Whitaker}, \& {Williams}}]{Hviding_2025}
{Hviding}, R.~E., {de Graaff}, A., {Miller}, T.~B., {et~al.} 2025, \bibinfo{title}{{RUBIES: A spectroscopic census of little red dots: All point sources with v-shaped continua have broad lines},} \aap, 702, A57, \dodoi{10.1051/0004-6361/202555816}

\bibitem[{K. {Inayoshi}(2025){Inayoshi}}]{Inayoshi_2025a}
{Inayoshi}, K. 2025, \bibinfo{title}{{Little Red Dots as the Very First Activity of Black Hole Growth},} \apjl, 988, L22, \dodoi{10.3847/2041-8213/adea66}

\bibitem[{K. {Inayoshi} \& L.~C. {Ho}(2025){Inayoshi} \& {Ho}}]{Inayoshi_Ho_2025}
{Inayoshi}, K., \& {Ho}, L.~C. 2025, \bibinfo{title}{{A Critical Evaluation of the Physical Nature of the Little Red Dots},} arXiv e-prints, arXiv:2512.03130, \dodoi{10.48550/arXiv.2512.03130}

\bibitem[{K. {Inayoshi} \& R. {Maiolino}(2025){Inayoshi} \& {Maiolino}}]{Inayoshi_Maiolino_2025}
{Inayoshi}, K., \& {Maiolino}, R. 2025, \bibinfo{title}{{Extremely Dense Gas around Little Red Dots and High-redshift Active Galactic Nuclei: A Nonstellar Origin of the Balmer Break and Absorption Features},} \apjl, 980, L27, \dodoi{10.3847/2041-8213/adaebd}

\bibitem[{K. {Inayoshi} {et~al.}(2025{\natexlab{a}}){Inayoshi}, {Murase}, \& {Kashiyama}}]{Inayoshi_2025d}
{Inayoshi}, K., {Murase}, K., \& {Kashiyama}, K. 2025{\natexlab{a}}, \bibinfo{title}{{Spectral Uniformity of Little Red Dots: A Natural Outcome of Coevolving Seed Black Holes and Nascent Starbursts},} arXiv e-prints, arXiv:2509.19422, \dodoi{10.48550/arXiv.2509.19422}

\bibitem[{K. {Inayoshi} {et~al.}(2025{\natexlab{b}}){Inayoshi}, {Shangguan}, {Chen}, {Ho}, \& {Haiman}}]{Inayoshi_2025b}
{Inayoshi}, K., {Shangguan}, J., {Chen}, X., {Ho}, L.~C., \& {Haiman}, Z. 2025{\natexlab{b}}, \bibinfo{title}{{The Emergence of Little Red Dots from Binary Massive Black Holes},} arXiv e-prints, arXiv:2505.05322, \dodoi{10.48550/arXiv.2505.05322}

\bibitem[{X. {Ji} {et~al.}(2025){Ji}, {Maiolino}, {{\"U}bler}, {Scholtz}, {D'Eugenio}, {Sun}, {Perna}, {Turner}, {Arribas}, {Bennett}, {Bunker}, {Carniani}, {Charlot}, {Cresci}, {Curti}, {Egami}, {Fabian}, {Inayoshi}, {Isobe}, {Jones}, {Juod{\v{z}}balis}, {Kumari}, {Lyu}, {Mazzolari}, {Parlanti}, {Robertson}, {Rodr{\'\i}guez Del Pino}, {Schneider}, {Sijacki}, {Tacchella}, {Trinca}, {Valiante}, {Venturi}, {Volonteri}, {Willott}, {Witten}, \& {Witstok}}]{Ji_2025}
{Ji}, X., {Maiolino}, R., {{\"U}bler}, H., {et~al.} 2025, \bibinfo{title}{{BlackTHUNDER -- A non-stellar Balmer break in a black hole-dominated little red dot at $z=7.04$},} arXiv e-prints, arXiv:2501.13082, \dodoi{10.48550/arXiv.2501.13082}

\bibitem[{X. {Ji} {et~al.}(2026){Ji}, {Pezzulli}, {D'Eugenio}, {Maiolino}, {Carniani}, {Tacchella}, {Jones}, {Smith}, {Witstok}, {Fabian}, {Geris}, {Harshan}, {Isobe}, {Ivey}, {Juod{\v{z}}balis}, {Pascalau}, {Scholtz}, \& {Witten}}]{Ji2026arXiv260403370J}
{Ji}, X., {Pezzulli}, G., {D'Eugenio}, F., {et~al.} 2026, \bibinfo{title}{{Holes in the BH$^\star$? AGN signatures in the FUV spectrum of a black-hole dominated Little Red Dot at $z=7.04$},} arXiv e-prints, arXiv:2604.03370, \dodoi{10.48550/arXiv.2604.03370}

\bibitem[{C.~M. {Johns-Krull}(2007){Johns-Krull}}]{Johns-Krull2007ApJ}
{Johns-Krull}, C.~M. 2007, \bibinfo{title}{{The Magnetic Fields of Classical T Tauri Stars},} \apj, 664, 975, \dodoi{10.1086/519017}

\bibitem[{P.~C. {Joss} {et~al.}(1973){Joss}, {Salpeter}, \& {Ostriker}}]{Joss1973ApJ}
{Joss}, P.~C., {Salpeter}, E.~E., \& {Ostriker}, J.~P. 1973, \bibinfo{title}{{On the ``Critical Luminosity'' in Stellar Interiors and Stellar Surface Boundary Conditions},} \apj, 181, 429, \dodoi{10.1086/152060}

\bibitem[{D. {Kido} {et~al.}(2025){Kido}, {Ioka}, {Hotokezaka}, {Inayoshi}, \& {Irwin}}]{Kido2025MNRAS}
{Kido}, D., {Ioka}, K., {Hotokezaka}, K., {Inayoshi}, K., \& {Irwin}, C.~M. 2025, \bibinfo{title}{{Black hole envelopes in Little Red Dots},} \mnras, 544, 3407, \dodoi{10.1093/mnras/staf1898}

\bibitem[{D.~D. {Kocevski} {et~al.}(2025){Kocevski}, {Finkelstein}, {Barro}, {Taylor}, {Calabr{\`o}}, {Laloux}, {Buchner}, {Trump}, {Leung}, {Yang}, {Dickinson}, {P{\'e}rez-Gonz{\'a}lez}, {Pacucci}, {Inayoshi}, {Somerville}, {McGrath}, {Akins}, {Bagley}, {Bowler}, {Bisigello}, {Carnall}, {Casey}, {Cheng}, {Cleri}, {Costantin}, {Cullen}, {Davis}, {Donnan}, {Dunlop}, {Ellis}, {Ferguson}, {Fujimoto}, {Fontana}, {Giavalisco}, {Grazian}, {Grogin}, {Hathi}, {Hirschmann}, {Huertas-Company}, {Holwerda}, {Illingworth}, {Juneau}, {Kartaltepe}, {Koekemoer}, {Li}, {Lucas}, {Magee}, {Mason}, {McLeod}, {McLure}, {Napolitano}, {Papovich}, {Pirzkal}, {Rodighiero}, {Santini}, {Wilkins}, \& {Yung}}]{Kocevski_2025}
{Kocevski}, D.~D., {Finkelstein}, S.~L., {Barro}, G., {et~al.} 2025, \bibinfo{title}{{The Rise of Faint, Red Active Galactic Nuclei at z > 4: A Sample of Little Red Dots in the JWST Extragalactic Legacy Fields},} \apj, 986, 126, \dodoi{10.3847/1538-4357/adbc7d}

\bibitem[{V. {Kokorev} {et~al.}(2025){Kokorev}, {Chisholm}, {Naidu}, {Fujimoto}, {Atek}, {Brammer}, {Finkelstein}, {Akins}, {Berg}, {Furtak}, {Fei}, {Hsiao}, {Labb{\'e}}, {Matthee}, {Mu{\~n}oz}, {Oesch}, {Pan}, {Rinaldi}, {Saldana-Lopez}, {Schaerer}, {Volonteri}, \& {Zitrin}}]{Kokorev_2025}
{Kokorev}, V., {Chisholm}, J., {Naidu}, R.~P., {et~al.} 2025, \bibinfo{title}{{The Deepest GLIMPSE of a Dense Gas Cocoon Enshrouding a Little Red Dot},} arXiv e-prints, arXiv:2511.07515, \dodoi{10.48550/arXiv.2511.07515}

\bibitem[{A.~K. {Kulkarni} \& M.~M. {Romanova}(2008){Kulkarni} \& {Romanova}}]{Kulkarni2008MNRAS}
{Kulkarni}, A.~K., \& {Romanova}, M.~M. 2008, \bibinfo{title}{{Accretion to magnetized stars through the Rayleigh-Taylor instability: global 3D simulations},} \mnras, 386, 673, \dodoi{10.1111/j.1365-2966.2008.13094.x}

\bibitem[{I. {Labb{\'e}} {et~al.}(2024){Labb{\'e}}, {Greene}, {Matthee}, {Treiber}, {Kokorev}, {Miller}, {Kramarenko}, {Setton}, {Ma}, {Goulding}, {Bezanson}, {Naidu}, {Williams}, {Atek}, {Brammer}, {Cutler}, {Chemerynska}, {Cloonan}, {Dayal}, {de Graaff}, {Fudamoto}, {Fujimoto}, {Furtak}, {Glazebrook}, {Heintz}, {Leja}, {Marchesini}, {Nanayakkara}, {Nelson}, {Oesch}, {Pan}, {Price}, {Shivaei}, {Sobral}, {Suess}, {van Dokkum}, {Wang}, {Weaver}, {Whitaker}, \& {Zitrin}}]{Labbe_2024b}
{Labb{\'e}}, I., {Greene}, J.~E., {Matthee}, J., {et~al.} 2024, \bibinfo{title}{{An unambiguous AGN and a Balmer break in an Ultraluminous Little Red Dot at z=4.47 from Ultradeep UNCOVER and All the Little Things Spectroscopy},} arXiv e-prints, arXiv:2412.04557, \dodoi{10.48550/arXiv.2412.04557}

\bibitem[{I. {Labb{\'e}} {et~al.}(2025){Labb{\'e}}, {Greene}, {Bezanson}, {Fujimoto}, {Furtak}, {Goulding}, {Matthee}, {Naidu}, {Oesch}, {Atek}, {Brammer}, {Chemerynska}, {Coe}, {Cutler}, {Dayal}, {Feldmann}, {Franx}, {Glazebrook}, {Leja}, {Maseda}, {Marchesini}, {Nanayakkara}, {Nelson}, {Pan}, {Papovich}, {Price}, {Suess}, {Wang}, {Weaver}, {Whitaker}, {Williams}, \& {Zitrin}}]{Labbe_2025}
{Labb{\'e}}, I., {Greene}, J.~E., {Bezanson}, R., {et~al.} 2025, \bibinfo{title}{{UNCOVER: Candidate Red Active Galactic Nuclei at 3 < z < 7 with JWST and ALMA},} \apj, 978, 92, \dodoi{10.3847/1538-4357/ad3551}

\bibitem[{F.~K. {Lamb} {et~al.}(1973){Lamb}, {Pethick}, \& {Pines}}]{Lamb1973ApJ}
{Lamb}, F.~K., {Pethick}, C.~J., \& {Pines}, D. 1973, \bibinfo{title}{{A Model for Compact X-Ray Sources: Accretion by Rotating Magnetic Stars},} \apj, 184, 271, \dodoi{10.1086/152325}

\bibitem[{X. {Lin} {et~al.}(2024){Lin}, {Wang}, {Fan}, {Cai}, {Champagne}, {Sun}, {Volonteri}, {Yang}, {Hennawi}, {Ba{\~n}ados}, {Barth}, {Eilers}, {Farina}, {Liu}, {Jin}, {Jun}, {Lupi}, {Kakiichi}, {Mazzucchelli}, {Onoue}, {Pan}, {Pizzati}, {Rojas-Ruiz}, {Schindler}, {Trakhtenbrot}, {Shen}, {Trebitsch}, {Zhuang}, {Endsley}, {Meyer}, {Li}, {Li}, {Pudoka}, {Tee}, {Wu}, \& {Zhang}}]{Lin_2024}
{Lin}, X., {Wang}, F., {Fan}, X., {et~al.} 2024, \bibinfo{title}{{A SPectroscopic Survey of Biased Halos In the Reionization Era (ASPIRE): Broad-line AGN at z = 4‑5 Revealed by JWST/NIRCam WFSS},} \apj, 974, 147, \dodoi{10.3847/1538-4357/ad6565}

\bibitem[{X. {Lin} {et~al.}(2026){Lin}, {Fan}, {Cai}, {Bian}, {Liu}, {Sun}, {Ma}, {Greene}, {Strauss}, {Green}, {Lyu}, {Champagne}, {Goulding}, {Inayoshi}, {Jin}, {Leung}, {Li}, {Liu}, {Liu}, {Mao}, {Pudoka}, {Tee}, {Wang}, {Wang}, {Wu}, {Yang}, {Zhang}, \& {Zhu}}]{Lin_2026}
{Lin}, X., {Fan}, X., {Cai}, Z., {et~al.} 2026, \bibinfo{title}{{The Discovery of Little Red Dots in the Local Universe: Signatures of Cool Gas Envelopes},} \apj, 997, 364, \dodoi{10.3847/1538-4357/ae2bdf}

\bibitem[{J.~L. {Linsky}(2017){Linsky}}]{Linsky2017ARA&A}
{Linsky}, J.~L. 2017, \bibinfo{title}{{Stellar Model Chromospheres and Spectroscopic Diagnostics},} \araa, 55, 159, \dodoi{10.1146/annurev-astro-091916-055327}

\bibitem[{H. {Liu} {et~al.}(2025){Liu}, {Jiang}, {Quataert}, {Greene}, \& {Ma}}]{H_Liu2025ApJ}
{Liu}, H., {Jiang}, Y.-F., {Quataert}, E., {Greene}, J.~E., \& {Ma}, Y. 2025, \bibinfo{title}{{The Balmer Break and Optical Continuum of Little Red Dots from Super-Eddington Accretion},} \apj, 994, 113, \dodoi{10.3847/1538-4357/ae0c19}

\bibitem[{J. {Lyu} {et~al.}(2024){Lyu}, {Alberts}, {Rieke}, {Shivaei}, {P{\'e}rez-Gonz{\'a}lez}, {Sun}, {Hainline}, {Baum}, {Bonaventura}, {Bunker}, {Egami}, {Eisenstein}, {Florian}, {Ji}, {Johnson}, {Morrison}, {Rieke}, {Robertson}, {Rujopakarn}, {Tacchella}, {Scholtz}, \& {Willmer}}]{Lyu_2024}
{Lyu}, J., {Alberts}, S., {Rieke}, G.~H., {et~al.} 2024, \bibinfo{title}{{Active Galactic Nuclei Selection and Demographics: A New Age with JWST/MIRI},} \apj, 966, 229, \dodoi{10.3847/1538-4357/ad3643}

\bibitem[{R. {Maiolino} {et~al.}(2024){Maiolino}, {Scholtz}, {Curtis-Lake}, {Carniani}, {Baker}, {de Graaff}, {Tacchella}, {{\"U}bler}, {D'Eugenio}, {Witstok}, {Curti}, {Arribas}, {Bunker}, {Charlot}, {Chevallard}, {Eisenstein}, {Egami}, {Ji}, {Jones}, {Lyu}, {Rawle}, {Robertson}, {Rujopakarn}, {Perna}, {Sun}, {Venturi}, {Williams}, \& {Willott}}]{Maiolino_2024_JADES}
{Maiolino}, R., {Scholtz}, J., {Curtis-Lake}, E., {et~al.} 2024, \bibinfo{title}{{JADES: The diverse population of infant black holes at 4 < z < 11: Merging, tiny, poor, but mighty},} \aap, 691, A145, \dodoi{10.1051/0004-6361/202347640}

\bibitem[{R. {Maiolino} {et~al.}(2025){Maiolino}, {Risaliti}, {Signorini}, {Trefoloni}, {Juod{\v{z}}balis}, {Scholtz}, {{\"U}bler}, {D'Eugenio}, {Carniani}, {Fabian}, {Ji}, {Mazzolari}, {Bertola}, {Brusa}, {Bunker}, {Charlot}, {Comastri}, {Cresci}, {DeCoursey}, {Egami}, {Fiore}, {Gilli}, {Perna}, {Tacchella}, \& {Venturi}}]{Maiolino_2025}
{Maiolino}, R., {Risaliti}, G., {Signorini}, M., {et~al.} 2025, \bibinfo{title}{{JWST meets Chandra: a large population of Compton thick, feedback-free, and intrinsically X-ray weak AGN, with a sprinkle of SNe},} \mnras, 538, 1921, \dodoi{10.1093/mnras/staf359}

\bibitem[{J. {Matthee} {et~al.}(2024){Matthee}, {Naidu}, {Brammer}, {Chisholm}, {Eilers}, {Goulding}, {Greene}, {Kashino}, {Labbe}, {Lilly}, {Mackenzie}, {Oesch}, {Weibel}, {Wuyts}, {Xiao}, {Bordoloi}, {Bouwens}, {van Dokkum}, {Illingworth}, {Kramarenko}, {Maseda}, {Mason}, {Meyer}, {Nelson}, {Reddy}, {Shivaei}, {Simcoe}, \& {Yue}}]{Matthee_2024}
{Matthee}, J., {Naidu}, R.~P., {Brammer}, G., {et~al.} 2024, \bibinfo{title}{{Little Red Dots: An Abundant Population of Faint Active Galactic Nuclei at z {\ensuremath{\sim}} 5 Revealed by the EIGER and FRESCO JWST Surveys},} \apj, 963, 129, \dodoi{10.3847/1538-4357/ad2345}

\bibitem[{J. {Matthee} {et~al.}(2026){Matthee}, {Torralba}, {Pezzulli}, {Naidu}, {Chisholm}, {Mascia}, {Greene}, {Ishikawa}, {Gronke}, {Wuyts}, {Bordoloi}, {Brammer}, {Chang}, {Eilers}, {de Graaff}, {Hviding}, {Iani}, {Illingworth}, {Kashino}, {Labbe}, {Ma}, {Maseda}, {Meyer}, {Nelson}, {Oesch}, \& {Xiao}}]{Matthee2026arXiv260317667M}
{Matthee}, J., {Torralba}, A., {Pezzulli}, G., {et~al.} 2026, \bibinfo{title}{{The Engine and its Flows: Little Red Dot spectra are shaped by the column densities of their gas envelopes},} arXiv e-prints, arXiv:2603.17667, \dodoi{10.48550/arXiv.2603.17667}

\bibitem[{M. {Mayer} \& W.~J. {Duschl}(2005){Mayer} \& {Duschl}}]{Mayer2005MNRAS}
{Mayer}, M., \& {Duschl}, W.~J. 2005, \bibinfo{title}{{Rosseland and Planck mean opacities for primordial matter},} \mnras, 358, 614, \dodoi{10.1111/j.1365-2966.2005.08826.x}

\bibitem[{R.~P. {Naidu} {et~al.}(2025){Naidu}, {Matthee}, {Katz}, {de Graaff}, {Oesch}, {Smith}, {Greene}, {Brammer}, {Weibel}, {Hviding}, {Chisholm}, {Labb\textbackslash'e}, {Simcoe}, {Witten}, {Atek}, {Baggen}, {Belli}, {Bezanson}, {Boogaard}, {Bose}, {Covelo-Paz}, {Dayal}, {Fudamoto}, {Furtak}, {Giovinazzo}, {Goulding}, {Gronke}, {Heintz}, {Hirschmann}, {Illingworth}, {Inoue}, {Johnson}, {Leja}, {Leonova}, {McConachie}, {Maseda}, {Natarajan}, {Nelson}, {Setton}, {Shivaei}, {Sobral}, {Stefanon}, {Tacchella}, {Toft}, {Torralba}, {van Dokkum}, {van der Wel}, {Volonteri}, {Walter}, {Wang}, \& {Watson}}]{Naidu_2025}
{Naidu}, R.~P., {Matthee}, J., {Katz}, H., {et~al.} 2025, \bibinfo{title}{{A ``Black Hole Star'' Reveals the Remarkable Gas-Enshrouded Hearts of the Little Red Dots},} arXiv e-prints, arXiv:2503.16596, \dodoi{10.48550/arXiv.2503.16596}

\bibitem[{Y. {Notsu} {et~al.}(2015){Notsu}, {Honda}, {Maehara}, {Notsu}, {Shibayama}, {Nogami}, \& {Shibata}}]{Notsu2015PASJ}
{Notsu}, Y., {Honda}, S., {Maehara}, H., {et~al.} 2015, \bibinfo{title}{{High dispersion spectroscopy of solar-type superflare stars. II. Stellar rotation, starspots, and chromospheric activities},} \pasj, 67, 33, \dodoi{10.1093/pasj/psv002}

\bibitem[{D.~E. {Osterbrock} \& G.~J. {Ferland}(2006){Osterbrock} \& {Ferland}}]{Osterbrock2006agna.book}
{Osterbrock}, D.~E., \& {Ferland}, G.~J. 2006, {Astrophysics of gaseous nebulae and active galactic nuclei}

\bibitem[{E.~N. {Parker}(1978){Parker}}]{Parker1978ApJ}
{Parker}, E.~N. 1978, \bibinfo{title}{{Hydraulic concentration of magnetic fields in the solar photosphere. VI. Adiabatic cooling and concentration in downdrafts.},} \apj, 221, 368, \dodoi{10.1086/156035}

\bibitem[{P.~G. {P{\'e}rez-Gonz{\'a}lez} {et~al.}(2024){P{\'e}rez-Gonz{\'a}lez}, {Barro}, {Rieke}, {Lyu}, {Rieke}, {Alberts}, {Williams}, {Hainline}, {Sun}, {Pusk{\'a}s}, {Annunziatella}, {Baker}, {Bunker}, {Egami}, {Ji}, {Johnson}, {Robertson}, {Rodr{\'\i}guez Del Pino}, {Rujopakarn}, {Shivaei}, {Tacchella}, {Willmer}, \& {Willott}}]{Perez-Gonzalez_2024}
{P{\'e}rez-Gonz{\'a}lez}, P.~G., {Barro}, G., {Rieke}, G.~H., {et~al.} 2024, \bibinfo{title}{{What Is the Nature of Little Red Dots and what Is Not, MIRI SMILES Edition},} \apj, 968, 4, \dodoi{10.3847/1538-4357/ad38bb}

\bibitem[{A.~A. {Pevtsov} {et~al.}(2003){Pevtsov}, {Fisher}, {Acton}, {Longcope}, {Johns-Krull}, {Kankelborg}, \& {Metcalf}}]{Pevtsov2003ApJ}
{Pevtsov}, A.~A., {Fisher}, G.~H., {Acton}, L.~W., {et~al.} 2003, \bibinfo{title}{{The Relationship Between X-Ray Radiance and Magnetic Flux},} \apj, 598, 1387, \dodoi{10.1086/378944}

\bibitem[{T. {Preibisch} {et~al.}(2005){Preibisch}, {Kim}, {Favata}, {Feigelson}, {Flaccomio}, {Getman}, {Micela}, {Sciortino}, {Stassun}, {Stelzer}, \& {Zinnecker}}]{Preibisch2005ApJS}
{Preibisch}, T., {Kim}, Y.-C., {Favata}, F., {et~al.} 2005, \bibinfo{title}{{The Origin of T Tauri X-Ray Emission: New Insights from the Chandra Orion Ultradeep Project},} \apjs, 160, 401, \dodoi{10.1086/432891}

\bibitem[{A. {Reiners}(2012){Reiners}}]{Reiners2012LRSP}
{Reiners}, A. 2012, \bibinfo{title}{{Observations of Cool-Star Magnetic Fields},} Living Reviews in Solar Physics, 9, 1, \dodoi{10.12942/lrsp-2012-1}

\bibitem[{A.~E. {Reines} {et~al.}(2013){Reines}, {Greene}, \& {Geha}}]{Reines2013ApJ}
{Reines}, A.~E., {Greene}, J.~E., \& {Geha}, M. 2013, \bibinfo{title}{{Dwarf Galaxies with Optical Signatures of Active Massive Black Holes},} \apj, 775, 116, \dodoi{10.1088/0004-637X/775/2/116}

\bibitem[{M.~M. {Romanova} \& S.~P. {Owocki}(2015){Romanova} \& {Owocki}}]{Romanova2015SSRv}
{Romanova}, M.~M., \& {Owocki}, S.~P. 2015, \bibinfo{title}{{Accretion, Outflows, and Winds of Magnetized Stars},} \ssr, 191, 339, \dodoi{10.1007/s11214-015-0200-9}

\bibitem[{M.~M. {Romanova} {et~al.}(2004){Romanova}, {Ustyugova}, {Koldoba}, \& {Lovelace}}]{Romanova2004ApJ}
{Romanova}, M.~M., {Ustyugova}, G.~V., {Koldoba}, A.~V., \& {Lovelace}, R.~V.~E. 2004, \bibinfo{title}{{Three-dimensional Simulations of Disk Accretion to an Inclined Dipole. II. Hot Spots and Variability},} \apj, 610, 920, \dodoi{10.1086/421867}

\bibitem[{R. {Rosner} {et~al.}(1978){Rosner}, {Tucker}, \& {Vaiana}}]{Rosner1978ApJ}
{Rosner}, R., {Tucker}, W.~H., \& {Vaiana}, G.~S. 1978, \bibinfo{title}{{Dynamics of the quiescent solar corona.},} \apj, 220, 643, \dodoi{10.1086/155949}

\bibitem[{V. {Rusakov} {et~al.}(2025){Rusakov}, {Watson}, {Nikopoulos}, {Brammer}, {Gottumukkala}, {Harvey}, {Heintz}, {Nielsen}, {Sim}, {Sneppen}, {Vijayan}, {Adams}, {Austin}, {Conselice}, {Goolsby}, {Toft}, \& {Witstok}}]{Rusakov_2025}
{Rusakov}, V., {Watson}, D., {Nikopoulos}, G.~P., {et~al.} 2025, \bibinfo{title}{{JWST's little red dots: an emerging population of young, low-mass AGN cocooned in dense ionized gas},} arXiv e-prints, arXiv:2503.16595, \dodoi{10.48550/arXiv.2503.16595}

\bibitem[{G.~G. {Sacco} {et~al.}(2010){Sacco}, {Orlando}, {Argiroffi}, {Maggio}, {Peres}, {Reale}, \& {Curran}}]{Sacco2010A&A}
{Sacco}, G.~G., {Orlando}, S., {Argiroffi}, C., {et~al.} 2010, \bibinfo{title}{{On the observability of T Tauri accretion shocks in the X-ray band},} \aap, 522, A55, \dodoi{10.1051/0004-6361/201014950}

\bibitem[{P.~N. {Safier}(1999){Safier}}]{Safier1999ApJ}
{Safier}, P.~N. 1999, \bibinfo{title}{{A Physical Limit to the Magnetic Fields of T Tauri Stars},} \apjl, 510, L127, \dodoi{10.1086/311807}

\bibitem[{J. {Scholtz} {et~al.}(2026){Scholtz}, {D'Eugenio}, {Maiolino}, {Brazzini}, {{\"U}bler}, {Ji}, {Perna}, {Sun}, {Brocchi}, {Carniani}, {Cresci}, {Ivey}, {Juod{\v{z}}balis}, {Marconi}, {Mazzolari}, {Risaliti}, \& {Trefoloni}}]{Scholtz2026arXiv260322277S}
{Scholtz}, J., {D'Eugenio}, F., {Maiolino}, R., {et~al.} 2026, \bibinfo{title}{{Little Red and Blue Dots: simply stratified Broad Line Regions},} arXiv e-prints, arXiv:2603.22277, \dodoi{10.48550/arXiv.2603.22277}

\bibitem[{C.~J. {Schrijver} {et~al.}(1989){Schrijver}, {Cote}, {Zwaan}, \& {Saar}}]{Schrijver1989ApJ}
{Schrijver}, C.~J., {Cote}, J., {Zwaan}, C., \& {Saar}, S.~H. 1989, \bibinfo{title}{{Relations between the Photospheric Magnetic Field and the Emission from the Outer Atmospheres of Cool Stars. I. The Solar CA II K Line Core Emission},} \apj, 337, 964, \dodoi{10.1086/167168}

\bibitem[{D.~J. {Setton} {et~al.}(2025){Setton}, {Greene}, {Spilker}, {Williams}, {Labb{\'e}}, {Ma}, {Wang}, {Whitaker}, {Leja}, {de Graaff}, {Alberts}, {Bezanson}, {Boogaard}, {Brammer}, {Cutler}, {Cleri}, {Cooper}, {Dayal}, {Fujimoto}, {Furtak}, {Goulding}, {Hirschmann}, {Kokorev}, {Maseda}, {McConachie}, {Matthee}, {Miller}, {Naidu}, {Oesch}, {Pan}, {Price}, {Suess}, {Weaver}, {Xiao}, {Zhang}, \& {Zitrin}}]{Setton_2025b}
{Setton}, D.~J., {Greene}, J.~E., {Spilker}, J.~S., {et~al.} 2025, \bibinfo{title}{{A Confirmed Deficit of Hot and Cold Dust Emission in the Most Luminous Little Red Dots},} \apjl, 991, L10, \dodoi{10.3847/2041-8213/ade78b}

\bibitem[{A. {Sneppen} {et~al.}(2026){Sneppen}, {Watson}, {Matthews}, {Nikopoulos}, {Allen}, {Brammer}, {Damgaard}, {Heintz}, {Knigge}, {Long}, {Rusakov}, {Sim}, \& {Witstok}}]{Sneppen2026arXiv260118864S}
{Sneppen}, A., {Watson}, D., {Matthews}, J.~H., {et~al.} 2026, \bibinfo{title}{{Inside the cocoon: a comprehensive explanation of the spectra of Little Red Dots},} arXiv e-prints, arXiv:2601.18864, \dodoi{10.48550/arXiv.2601.18864}

\bibitem[{L. {Spitzer}(1956){Spitzer}}]{Spitzer1956}
{Spitzer}, L. 1956, {Physics of Fully Ionized Gases}

\bibitem[{H.~C. {Spruit}(1976){Spruit}}]{Spruit1976SoPh}
{Spruit}, H.~C. 1976, \bibinfo{title}{{Pressure equilibrium and energy balance of small photospheric fluxtubes.},} \solphys, 50, 269, \dodoi{10.1007/BF00155292}

\bibitem[{S.~W. {Stahler} {et~al.}(1986){Stahler}, {Palla}, \& {Salpeter}}]{Stahler1986ApJ}
{Stahler}, S.~W., {Palla}, F., \& {Salpeter}, E.~E. 1986, \bibinfo{title}{{Primordial Stellar Evolution: The Protostar Phase},} \apj, 302, 590, \dodoi{10.1086/164018}

\bibitem[{W.~Q. {Sun} {et~al.}(2026){Sun}, {Naidu}, {Matthee}, {de Graaff}, {Chisholm}, {Greene}, {Oesch}, {Torralba}, {Hviding}, {Brammer}, {Simcoe}, {Bose}, {Bouwens}, {Dayal}, {Eilers}, {Fei}, {Furtak}, {Gottumukkala}, {Goulding}, {Heintz}, {Hirschmann}, {Kokorev}, {Leja}, {Liu}, {Natarajan}, {Santarelli}, {Setton}, {Smith}, {Tacchella}, {Volonteri}, {Walter}, {Weibel}, \& {Williams}}]{W.Sun_2026}
{Sun}, W.~Q., {Naidu}, R.~P., {Matthee}, J., {et~al.} 2026, \bibinfo{title}{{Little Red Dot $-$ Host Galaxy $=$ Black Hole Star: A Gas-Enshrouded Heart at the Center of Every Little Red Dot},} arXiv e-prints, arXiv:2601.20929, \dodoi{10.48550/arXiv.2601.20929}

\bibitem[{S. {Takasao} {et~al.}(2025){Takasao}, {Kunitomo}, {Suzuki}, {Iwasaki}, \& {Tomida}}]{Takasao2025ApJ_a}
{Takasao}, S., {Kunitomo}, M., {Suzuki}, T.~K., {Iwasaki}, K., \& {Tomida}, K. 2025, \bibinfo{title}{{Spin-down of Solar-mass Protostars in Magnetospheric Accretion Paradigm},} \apj, 980, 111, \dodoi{10.3847/1538-4357/ada364}

\bibitem[{S. {Takasao} {et~al.}(2020){Takasao}, {Mitsuishi}, {Shimura}, {Yoshida}, {Kunitomo}, {Tanaka}, \& {Ishihara}}]{Takasao2020ApJ}
{Takasao}, S., {Mitsuishi}, I., {Shimura}, T., {et~al.} 2020, \bibinfo{title}{{Investigation of Coronal Properties of X-Ray Bright G-dwarf Stars Based on the Solar Surface Magnetic Field-Corona Relationship},} \apj, 901, 70, \dodoi{10.3847/1538-4357/abad34}

\bibitem[{S. {Takasao} {et~al.}(2022){Takasao}, {Tomida}, {Iwasaki}, \& {Suzuki}}]{Takasao2022ApJ}
{Takasao}, S., {Tomida}, K., {Iwasaki}, K., \& {Suzuki}, T.~K. 2022, \bibinfo{title}{{Three-dimensional Simulations of Magnetospheric Accretion in a T Tauri Star: Accretion and Wind Structures Just Around the Star},} \apj, 941, 73, \dodoi{10.3847/1538-4357/ac9eb1}

\bibitem[{M. {Tang} {et~al.}(2025){Tang}, {Stark}, {Plat}, {Feltre}, {Katz}, {Senchyna}, {Mason}, {Whitler}, {Chen}, \& {Topping}}]{Tang2025ApJ}
{Tang}, M., {Stark}, D.~P., {Plat}, A., {et~al.} 2025, \bibinfo{title}{{JWST/NIRSpec Observations of High-ionization Emission Lines in Galaxies at High Redshift},} \apj, 991, 217, \dodoi{10.3847/1538-4357/adfd57}

\bibitem[{M. {Tang} {et~al.}(2026){Tang}, {Stark}, {Mason}, {Chen}, {Katz}, {Gronke}, {Furtak}, {Chang}, {Matthee}, {Whitler}, {Zitrin}, {Endsley}, {Gelli}, {Roychowdhury}, {Senchyna}, {Topping}, \& {Zhang}}]{Tang2026arXiv260403563T}
{Tang}, M., {Stark}, D.~P., {Mason}, C.~A., {et~al.} 2026, \bibinfo{title}{{SPURS: Evidence for Clumpy Neutral Envelopes and Ionized IGM Surrounding Little Red Dots in Abell 2744 from Ultra-Deep Rest-UV Spectroscopy},} arXiv e-prints, arXiv:2604.03563, \dodoi{10.48550/arXiv.2604.03563}

\bibitem[{A.~J. {Taylor} {et~al.}(2025){Taylor}, {Kokorev}, {Kocevski}, {Akins}, {Cullen}, {Dickinson}, {Finkelstein}, {Arrabal Haro}, {Bromm}, {Giavalisco}, {Inayoshi}, {Juneau}, {Leung}, {P{\'e}rez-Gonz{\'a}lez}, {Somerville}, {Trump}, {Amor{\'\i}n}, {Barro}, {Burgarella}, {Brooks}, {Carnall}, {Casey}, {Cheng}, {Chisholm}, {Chworowsky}, {Davis}, {Donnan}, {Dunlop}, {Ellis}, {Fern{\'a}ndez}, {Fujimoto}, {Grogin}, {Gupta}, {Hathi}, {Jung}, {Hirschmann}, {Kartaltepe}, {Koekemoer}, {Larson}, {Leung}, {Llerena}, {Lucas}, {McLeod}, {McLure}, {Napolitano}, {Papovich}, {Stanton}, {Tripodi}, {Wang}, {Wilkins}, {Yung}, \& {Zavala}}]{Taylor_2025b}
{Taylor}, A.~J., {Kokorev}, V., {Kocevski}, D.~D., {et~al.} 2025, \bibinfo{title}{{CAPERS-LRD-z9: A Gas-enshrouded Little Red Dot Hosting a Broad-line Active Galactic Nucleus at z = 9.288},} \apjl, 989, L7, \dodoi{10.3847/2041-8213/ade789}

\bibitem[{B. {Tessore} {et~al.}(2017){Tessore}, {L{\`e}bre}, {Morin}, {Mathias}, {Josselin}, \& {Auri{\`e}re}}]{Tessore2017A&A}
{Tessore}, B., {L{\`e}bre}, A., {Morin}, J., {et~al.} 2017, \bibinfo{title}{{Measuring surface magnetic fields of red supergiant stars},} \aap, 603, A129, \dodoi{10.1051/0004-6361/201730473}

\bibitem[{O.~D. {Toropina} {et~al.}(2005){Toropina}, {Romanova}, {Toropin}, \& {Lovelace}}]{Toropina2005MmSAI}
{Toropina}, O.~D., {Romanova}, M.~M., {Toropin}, Y.~M., \& {Lovelace}, R.~V.~E. 2005, \bibinfo{title}{{Spherical Accretion to a Magnetized Neutron Star in the ``Propeller'' Regime.},} \memsai, 76, 508

\bibitem[{A. {Ud-Doula} {et~al.}(2008){Ud-Doula}, {Owocki}, \& {Townsend}}]{ud-Doula2008MNRAS}
{Ud-Doula}, A., {Owocki}, S.~P., \& {Townsend}, R. H.~D. 2008, \bibinfo{title}{{Dynamical simulations of magnetically channelled line-driven stellar winds - II. The effects of field-aligned rotation},} \mnras, 385, 97, \dodoi{10.1111/j.1365-2966.2008.12840.x}

\bibitem[{H. {Umeda} {et~al.}(2026){Umeda}, {Inayoshi}, {Harikane}, \& {Murase}}]{Umeda2026ApJ}
{Umeda}, H., {Inayoshi}, K., {Harikane}, Y., \& {Murase}, K. 2026, \bibinfo{title}{{A Black Hole Envelope Interpretation for Cosmological Demographics of Little Red Dots},} \apj, 999, 183, \dodoi{10.3847/1538-4357/ae4101}

\bibitem[{W.~H.~T. {Vlemmings} {et~al.}(2005){Vlemmings}, {van Langevelde}, \& {Diamond}}]{Vlemmings2005A&A}
{Vlemmings}, W.~H.~T., {van Langevelde}, H.~J., \& {Diamond}, P.~J. 2005, \bibinfo{title}{{The magnetic field around late-type stars revealed by the circumstellar H\_2O masers},} \aap, 434, 1029, \dodoi{10.1051/0004-6361:20042488}

\bibitem[{B. {Wang} {et~al.}(2025){Wang}, {Leja}, {Katz}, {Inayoshi}, {Cleri}, {de Graaff}, {Hviding}, {van Dokkum}, {Greene}, {Labb{\'e}}, {Matthee}, {McConachie}, {Naidu}, \& {Nelson}}]{Wang_2025b}
{Wang}, B., {Leja}, J., {Katz}, H., {et~al.} 2025, \bibinfo{title}{{The Missing Hard Photons of Little Red Dots: Their Incident Ionizing Spectra Resemble Massive Stars},} arXiv e-prints, arXiv:2508.18358, \dodoi{10.48550/arXiv.2508.18358}

\end{thebibliography}
\bibliographystyle{aasjournalv7}



\end{document}